\documentclass[a4paper,11pt]{article}
\pdfoutput=1

\usepackage{jheppub}

\usepackage[T1]{fontenc}
\usepackage{subcaption}
\usepackage{caption}
\usepackage{slashed}
\usepackage{physics}
\usepackage{wrapfig}

\title{Thermal QCD in a non-uniform magnetic background}

\author[a]{B. B. Brandt,}
\author[b]{F. Cuteri,}
\author[a]{G. Endr\H{o}di,}
\author[a]{G. Mark\'o,}
\author[a]{L. Sandbote,}
\author[a]{A. D. M. Valois}

\affiliation[a]{Universität Bielefeld,\\
  Universitätsstraße 25, 33615 Bielefeld, Germany}
\affiliation[b]{Institute for Theoretical Physics, Goethe University,\\
 Max-von-Laue-Straße 1, 60438 Frankfurt, Germany}

\emailAdd{brandt@physik.uni-bielefeld.de}
\emailAdd{cuteri@itp.uni-frankfurt.de}
\emailAdd{endrodi@physik.uni-bielefeld.de}
\emailAdd{gmarko@physik.uni-bielefeld.de}
\emailAdd{leon.sandbote@uni-bielefeld.de}
\emailAdd{dvalois@physik.uni-bielefeld.de}

\newcommand{\Z}{\mathcal{Z}}
\newcommand{\ave}[1]{\left\langle#1\right\rangle}

\abstract{
Off-central heavy-ion collisions are known to feature magnetic fields with 
magnitudes and characteristic gradients corresponding to the scale of the strong interactions. In this work, we employ equilibrium lattice simulations of the underlying theory, QCD, involving similar inhomogeneous magnetic field profiles to 
achieve a better understanding of this system.
We simulate three flavors of dynamical staggered quarks with physical masses 
at a range of magnetic fields and temperatures, and extrapolate the results to the continuum limit. Analyzing the impact of the field on the 
quark condensate and the Polyakov loop, we find non-trivial spatial features that render the QCD medium qualitatively different as in the homogeneous setup, especially at temperatures around the transition. In addition, we construct leading-order chiral perturbation theory for the inhomogeneous background and compare its prediction to our lattice results at low temperature.
Our findings will be useful to benchmark effective theories and low-energy models of QCD for a better description of peripheral heavy-ion collisions.}

\begin{document} 
\maketitle
\flushbottom

\section{Introduction}\label{sec:introduction}
Strong magnetic fields are known to appear in several physical systems around us in the Universe. Magnetars, for instance, are strongly magnetized neutron stars with magnetic fields up to $10^{15}$ G ($\sqrt{eB}\sim1$ MeV) in their interior~\cite{duncan1992formation}. During mergers of such compact objects considerably high temperatures are also achieved~\cite{Most:2018eaw}. Experiments of non-central heavy-ion collisions can produce transient fields of strength $10^{18}$ - $10^{19}$ G ($\sqrt{eB}\sim0.1$ - $0.5$ GeV) at RHIC and LHC energies, respectively~\cite{skokov2009estimate}. Furthermore, cosmological models predict even stronger fields emerging during the electroweak phase of the early Universe, with magnitudes as high as $10^{20}$ G ($\sqrt{eB}\sim1.5$ GeV)~\cite{vachaspati1991magnetic}. In turn, strong magnetic fields might also be viewed as valuable probes to study the intricate structure of the non-perturbative QCD medium. Since the strength of the above fields is comparable to the QCD energy scale, the interplay of strong and electromagnetic forces must be taken into account to reach a better understanding of how quarks and gluons behave in the systems mentioned above. 

In this work, we will focus on the impact of magnetic fields in the context of heavy-ion collisions. The case of strong uniform fields has been widely studied both numerically on the lattice and analytically via QCD models. For a recent review on these approaches, see e.g.~Ref.~\cite{andersen2016phase}. However, in heavy-ion collision experiments, the generated fields deviate significantly from the uniform case. In addition, since the magnetic field also changes rapidly with time (see, e.g.\ the review~\cite{Huang:2015oca}), it also induces non-uniform time-dependent electric fields. These might play a role in the subsequent dynamics of the byproducts of the collision and the behavior of the quark-gluon plasma.
In particular, event-by-event simulations of heavy-ion collisions predicted nontrivial profiles for both the electric and magnetic field components \cite{voronyuk2011electromagnetic,deng2012event}, with potential impact for particle flow~\cite{Pang:2016yuh}.

The explicit time dependence and the presence of (real) electric 
fields cannot be implemented using standard lattice QCD simulations in
equilibrium due to the infamous complex action problem.\footnote{Note the recent approach to include electric fields on the lattice at nonzero temperature within a Taylor-expansion~\cite{Endrodi:2022wym,Endrodi:2023wwf}.}
However, well established simulation algorithms can be employed 
to include inhomogeneous magnetic fields. This way we can take one 
step towards a better description of the heavy-ion collision scenario
that helps capture the correct physics in peripheral events. 
In this paper we consider, for the first time in lattice QCD simulations, a non-uniform background magnetic field $B(x)$ modulated in one spatial direction. Our choice $\cosh^{-2}(x/\epsilon)$ for the profile is motivated by the inhomogeneities observed in the above mentioned model simulations of heavy-ion collisions, as well as due to the possibility of an analytical treatment of the free Dirac operator in this case~\cite{Dunne:2004nc,cao2018chiral}.
We note that our approach can be generalized to introduce modulations in the other spatial directions, too. 
Our preliminary results have already been discussed in~\cite{MarquesValois:2021kvf}.

The magnetic field is known to lower the QCD transition temperature, thereby preferring deconfinement and chiral symmetry restoration~\cite{bali2012qcd}. The presence of an inhomogeneous magnetic field therefore raises the intriguing possibility of an inhomogeneous medium, where regions exposed to strong and weak $B$ resemble different phases of the theory. 
If the magnetic field is so strong that the homogeneous setup exhibits a first-order phase transition (as recently observed in Ref.~\cite{DElia:2021yvk}), the different regions of the inhomogeneous system might even become separated by sharp boundaries, leading to novel physical consequences.
Furthermore, the characteristic features for deconfinement and for chiral symmetry restoration may be affected differently by the magnetic field, making the inhomogeneous field an effective novel probe of the QCD medium.

This paper is organized as follows: in Sec.~\ref{sec:lattice_intro} we discuss the basic aspects of magnetic fields on the lattice, reviewing the flux quantization for homogeneous fields, generalizing it to the inhomogeneous case and discussing valence and sea effects. Sec.~\ref{sec:simulation_setup} summarizes the details of our simulations, followed by Sec.~\ref{sec:results}, where we present our results for the local observables. Finally, we summarize our findings in Sec.~\ref{sec:conclusions}. In the appendices we give details on the implementation of the inhomogeneous magnetic field and the eigenvalues of free fermions and bosons on such backgrounds. The latter is relevant for the construction of chiral perturbation theory for this setup, which we perform here for the first time.
\section{Magnetic fields in QCD and on the lattice}
\label{sec:lattice_intro}

\subsection{Homogeneous magnetic fields}
\label{sec:hom}
In this section we will review how homogeneous magnetic fields are implemented on the lattice, how key observables are defined and discuss their behaviour to set the stage for the inhomogeneous case. Throughout we will consider lattice geometries with $N_\mu$ points in the $\mu$ direction and a lattice spacing $a$. The spatial sizes, the volume and the temperature are given by $L_i=N_ia$, $V=L_xL_yL_z$ and $T=(N_ta)^{-1}$. The coordinates run over a symmetric range, i.e.\ $-L_x/2\le x <L_x/2$. Our boundary conditions are periodic in space and (anti-)periodic for bosonic (fermionic) fields in imaginary time. 

In order to implement the magnetic field on the lattice, besides the non-Abelian $\mathrm{SU}(3)$ links corresponding to the gluon fields in QCD, we must also introduce Abelian links $u_{\mu}=e^{ia q A_\mu}\in \mathrm{U}(1)$ representing the electromagnetic field $A_\mu$ for a quark with electric charge $q$. For a homogeneous field $B$ pointing in the $z$ direction, realized by the gauge $A_y=Bx$, a simple choice of links would be $u_y = e^{iaqBx}$ and $u_x=u_z=u_t=1$. However, this prescription is not periodic: $u_y(x=L_x/2) = e^{iaqBL_x/2} \neq u_y(x=-L_x/2)$. For lattice simulations it is convenient to have the $\mathrm{U}(1)$ links satisfy periodic boundary conditions. Therefore, we perform a gauge transformation and derive an equivalent prescription for the links in the uniform case
\begin{align}
\label{eq:hom_links}
u_{x}(x,y,z,t) &= 
    \left\{
        \begin{array}{ll}
        e^{-iqB L_x (y+L_y/2)} \qquad & \mbox{if } x = L_x/2-a\,, \nonumber\\
        1 & \mbox{otherwise}\,,
        \end{array}
    \right. \\
u_{y}(x,y,z,t) &= e^{iaqBx}\,, \\
u_z(x,y,z,t) &= 1\,, \nonumber\\
u_t(x,y,z,t) &= 1\,. \nonumber
\end{align}
The periodicity of the lattice imposes that the magnetic flux has to be quantized according to~\cite{tHooft:1979rtg}
\begin{equation}
qB = \frac{2\pi N_b}{L_xL_y},\hspace{1cm} N_b\in\mathbb{Z}.
\end{equation}

The QCD partition function in the presence of a uniform background magnetic field using the rooted staggered formulation of fermions is
\begin{equation}
    \Z(B) = \int \mathcal{D}U\hspace{0.1cm} e^{-S_g} \prod_{f=u,d,s}\left[\det(\slashed{D}(U,q_fB)+m_f)\right]^{1/4}\,,
\end{equation}
where the magnetic field dependence enters through the Dirac operator $\slashed{D}$ of each flavor. The gluon action $S_g$ will be specified below in Sec.~\ref{sec:simulation_setup}. The Abelian links entering the Dirac operator of a certain flavor depend on the combination $q_f B$, see  \eqref{eq:hom_links}, which we indicated explicitly.
We will concentrate on two observables, the quark condensate and the Polyakov loop. The former is defined via the derivative of $\log \Z(B)$ with respect to $m_f$,
\begin{equation}
    \ave{\bar{\psi}\psi_f}_{B,T} = \frac{T}{4V}\frac{1}{\Z(B)}\int \mathcal{D}U\hspace{0.1cm} e^{-S_g}\prod_{f'}\left[\det(\slashed{D}(U,q_{f'}B)+m_{f'})\right]^{1/4}\Tr\left[(\slashed{D}(U,q_{f}B)+m_f)^{-1}\right]\,,\label{eq:chiral-condensate-hom}
\end{equation}
with $\Tr$ standing for a trace in both Dirac and color spaces as well as a summation over the full lattice. In turn, the Polyakov loop is defined as
\begin{equation}
\ave{P}_{B,T} = \frac{1}{V}\frac{1}{\Z(B)}\int \mathcal{D}U\hspace{0.1cm} e^{-S_g}\prod_{f}\left[\det(\slashed{D}(U,q_fB)+m_f)\right]^{1/4}\sum_{x,y,z}\Re\tr[\prod_{t}U_t(x,y,z)]\,,\label{eq:polyakov-loop-hom}
\end{equation}
and the $\tr$ stands for a trace only in color space.

Both of these observables need renormalization. The quark condensate, as defined in Eq.~\eqref{eq:chiral-condensate-hom}, contains an additive and a multiplicative ultraviolet divergence, which are, however, independent of the temperature and the magnetic field. Following Ref.~\cite{Bali:2012zg}, we renormalize it by defining
\begin{equation}
\Sigma_f(B,T) = \frac{2m_{f}}{m_{\pi}^2 F^2}\qty[\ave{\bar{\psi}\psi_f}_{B,T} - \ave{\bar{\psi}\psi_f}_{B=T=0}]+1\,.
\label{eq:Sigmadef}
\end{equation}
Here, the subtraction of the zero-field and zero-temperature condensate eliminates the additive divergence, whereas the multiplication by $m_{f}$ cancels the multiplicative one. The normalization factor involving the physical pion mass $m_{\pi}=135$ MeV and the chiral limit of the pion decay constant $F=86\textmd{ MeV}$ is for convenience. The combination~\eqref{eq:Sigmadef} is unity in the vacuum and, according to the Gell-Mann-Oakes-Renner relation, it approaches zero for high temperatures. We define the condensate for the light quarks as $\Sigma\equiv(\Sigma_u+\Sigma_d)/2$, which we use as the approximate order parameter in the chiral sector.

The Polyakov loop also contains an ultraviolet divergence, which can be eliminated via~\cite{Borsanyi:2012uq}
\begin{equation}
P_R(B,T)=W(a,T)\cdot\expval{P}_{B,T}\,,
\end{equation}
where the multiplicative factor $W(a,T)$ is 
independent of the magnetic field and has been measured for our action in Ref.~\cite{Bruckmann:2013oba}.

\subsection{Inhomogeneous magnetic fields}
\label{sec:mag_field}

\begin{figure}[!h]
\centering
    \includegraphics[width=0.5\textwidth]{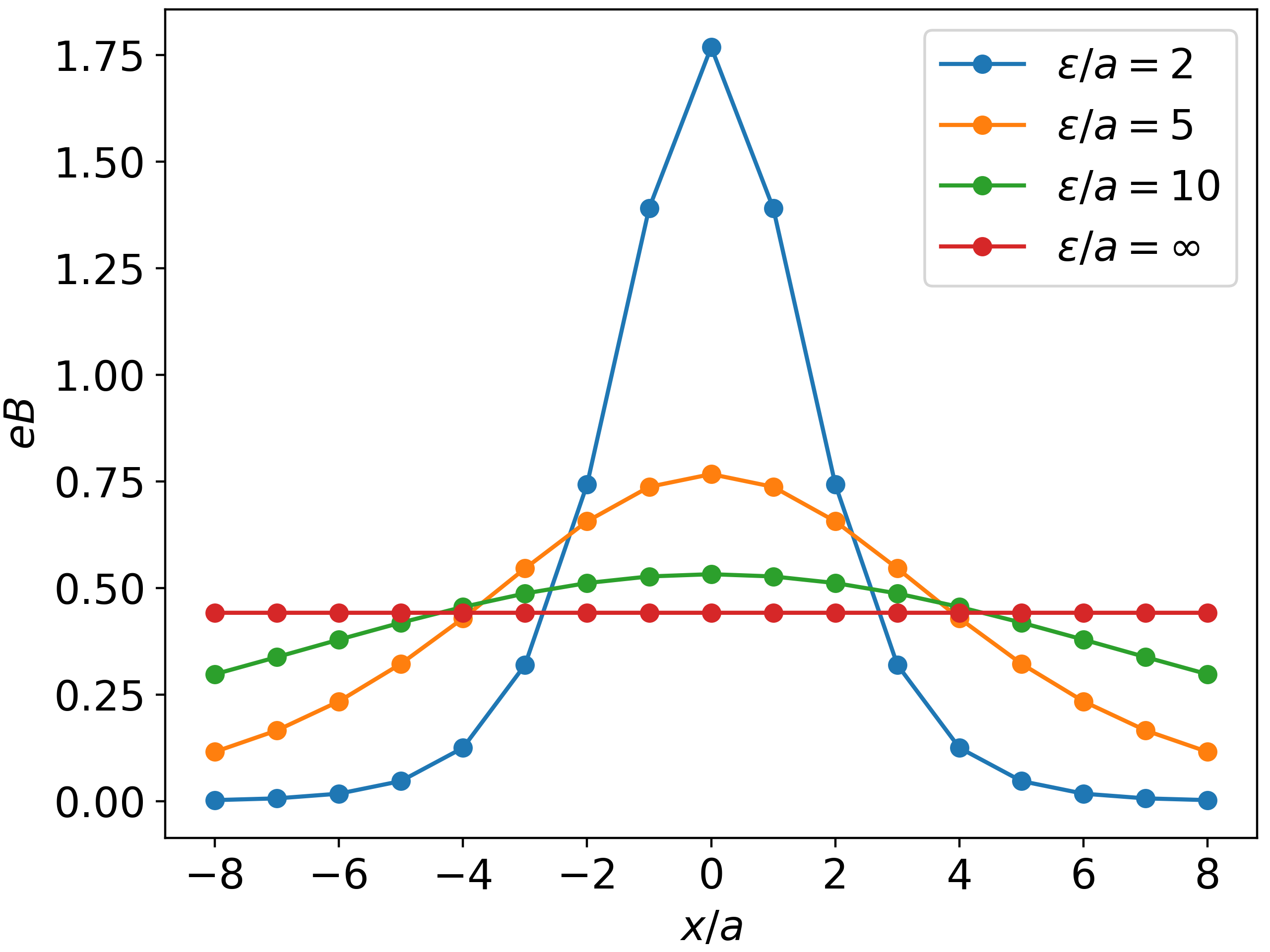}
    \caption{The profile~\protect\eqref{eq:inv_cosh_profile} of the magnetic field in lattice units for different values of the parameter $\epsilon$ on an $N_x=16$ lattice. For $\epsilon\to\infty$ the homogeneous profile is recovered.}
    \label{fig:Bprofile}
\end{figure}

In order to implement inhomogeneous fields on the lattice we need to extend the procedure described in the beginning of Sec.~\ref{sec:hom}. The details of the generalization are relegated to App.~\ref{app:u1links}, here we only summarize the main results concerning our choice of inhomogeneous field profile. We consider a magnetic field pointing in the $z$ direction with the particular form
\begin{equation}
\Vec{B}(x) = B \cosh^{-2}\qty(\frac{x}{\epsilon})\,\hat{z}\,, %
\label{eq:inv_cosh_profile}
\end{equation}
where $\epsilon$ is the width of the profile, centered in the middle of the lattice and $B$ is the magnitude of the field at the center.  Fig.~\ref{fig:Bprofile} illustrates the shape of the profile for different values of $\epsilon$. 
Similarly to the uniform case, the flux of the inhomogeneous field is also quantized,
\begin{equation}
qB = \frac{\pi N_b}{L_y\epsilon\tanh(L_x/2\epsilon)},\hspace{1cm} N_b\in\mathbb{Z}.
\label{eq:B-quantization-rule}
\end{equation}
allowing us to parameterize the field with the integer $N_b$ and the continuous variable $\epsilon$. As Fig.~\ref{fig:lattice_cartoon} shows, the homogeneous field is recovered by keeping $N_b$ constant and approaching $\epsilon\to\infty$.
The prescription for the links in this case is (see App.~\ref{app:u1links}),
\begin{align}
u_{x}(x,y,z,t) &= 
    \left\{
        \begin{array}{ll}
        e^{-i2\pi N_b (y/L_y + 1/2)} \qquad & \mbox{if } x = L_x/2-a\,, \nonumber \\
        1 & \mbox{otherwise}\,,
        \end{array}
    \right. \\
u_{y}(x,y,z,t) &= e^{iqB\epsilon a\qty[\tanh(\frac{x}{\epsilon}) + \tanh(\frac{L_x}{2\epsilon})]}\,, \label{eq:links1} \\
u_z(x,y,z,t) &= 1\,, \nonumber\\
u_t(x,y,z,t) &= 1\,. \nonumber
\end{align}
It is interesting to note that only the $\mathrm{U}(1)$ links in the bulk of 
the volume are sensitive to the particular spatial dependence of the magnetic field -- the $u_x$ links originating from the gauge transformation required for periodicity only depend on the flux quantum $N_b$. Consequently, the twisted $u_x$ links for the inhomogeneous~\eqref{eq:links1} and homogeneous ~\eqref{eq:hom_links} profiles are identical.
Notice  moreover that even though $B(x)$ is periodic, its derivatives differ at $x=-L_x/2$ and $x=L_x/2$. This is, however unproblematic as long as $\epsilon/L_x$ is sufficiently small, i.e.\ the volume is large compared to the profile width. We point out here, that from now on, whenever we refer to the value of the field, we mean the amplitude $B$, unless otherwise explicitly noted.
We also extend our list of observables to probe the effects of the inhomogeneous $B$-field.  Specifically, we define local (at least in one spatial dimension) versions of Eqs.~\eqref{eq:chiral-condensate-hom} and~\eqref{eq:polyakov-loop-hom}. Then, the local quark condensate for a flavor $f$ at finite $B$ is given by 
\begin{equation}
\label{eq:local_cond_bare}
\ave{\bar{\psi}\psi_f(x)}_{B,T} = \frac{T}{4L_yL_z}\frac{1}{\Z(B)}\int \mathcal{D}U\hspace{0.1cm} e^{-S_g}\prod_{f'}\left[\det(\slashed{D}(U,q_{f'}B)+m_{f'})\right]^{1/4}\widehat\Tr\left[(\slashed{D}(U,q_{f}B)+m_f)^{-1}\right]\,,
\end{equation}
where $\widehat\Tr$ stands for a trace in Dirac and color space and a summation over the $y$, $z$ and $t$ coordinates, retaining the $x$-dependence of the local quark condensate. With a similar mindset, the local Polyakov loop is defined as
\begin{equation}
\ave{P(x)}_{B,T} = \frac{1}{L_yL_z}\frac{1}{\Z(B)}\int \mathcal{D}U\hspace{0.1cm} e^{-S_g}\prod_{f}\left[\det(\slashed{D}(U,q_fB)+m_f)\right]^{1/4}\sum_{y,z}\Re\tr[\prod_{t}U_t(x,y,z,t)]\,.\label{eq:local_pol}
\end{equation} 

The renormalization of the local objects is carried out by the same procedure as for the fully volume averaged quantities of Sec.~\ref{sec:hom}, as the ultraviolet divergences do not depend on the coordinate. Therefore the renormalized local condensate, generalizing~\eqref{eq:Sigmadef}, is defined as
\begin{equation}
\Sigma_f(x,B,T) = \frac{2m_{f}}{m_{\pi}^2F^2}\qty[\ave{\bar{\psi}\psi_f(x)}_{B,T} - \ave{\bar{\psi}\psi_f}_{B=T=0}]+1\,,
\label{eq:Sigmafdef1}
\end{equation}
and the average light quark condensate is denoted by $\Sigma(x)\equiv(\Sigma_u(x)+\Sigma_d(x))/2$. We can also define the modification of the local chiral condensate relatively to the zero-field one as
\begin{equation}
\Delta\Sigma(x,B,T) = \Sigma(x,B,T)-\Sigma(x,0,T)
\label{eq:chiral_cond_shift}
\end{equation}
which will be useful for our analysis later on.
Finally, the renormalized local Polyakov loop is
\begin{equation}
    \label{eq:deltaP}
    P_R(x,B,T)=W(a,T)\cdot \expval{P(x)}_{B,T}\,.
\end{equation}

\subsection{Valence and sea effects}
\label{sec:valsea}

As we will see, the magnetic field has a nontrivial impact on the observables that can be best understood in terms of a separation of valence and sea effects.
Let us first summarize what we know about these in the uniform magnetic field case.
For the condensate, the magnetic field dependence appears in two ways~\cite{d2011chiral}. First, in the trace in Eq.~\eqref{eq:chiral-condensate-hom}, which represents the direct coupling between the electrically charged quarks and $B$ and is called the valence effect. This coupling tends to enhance the condensate and this enhancement is often referred to as {\it magnetic catalysis}. 
Magnetic catalysis can be understood in various different ways: for strong magnetic fields it emerges due to the dimensional reduction of the system~\cite{Shovkovy:2012zn}, while at low $B$ it is related to the positivity of the QED $\beta$-function~\cite{Endrodi:2013cs,Bali:2013txa}. Either way, magnetic catalysis is deeply connected to the proliferation of 
low Dirac eigenvalues, which has been demonstrated explicitly for full QCD~\cite{Bruckmann:2017pft}. 

Second, the magnetic field also has an indirect effect on the distribution of gluonic configurations via sea quark loops, i.e.\ via the determinant in Eq.~\eqref{eq:chiral-condensate-hom}. This is the so-called sea effect. Its main impact for the gluon fields is to enhance the Polyakov loop in the transition region~\cite{Bruckmann:2013oba}, which, in turn, correlates with the suppression of the condensate.
The sea effect is found to dominate the valence effect in the transition region, leading to an overall reduction of $\langle\bar\psi\psi\rangle$ for the light quarks due to $B$, dubbed {\it inverse magnetic catalysis}. In turn, away from the transition temperature, magnetic catalysis prevails. Altogether, the effect of the magnetic field is to reduce the transition temperature, defined via the inflection point of the condensate. The picture that we just sketched holds for physical quark masses and the inverse magnetic catalysis phenomenon disappears for heavier quarks. However, the reduction of $T_c(B)$ was found to persist for all quark masses~\cite{d2018qcd,endrHodi2019magnetic}.

In our inhomogenous setup, the condensate~\eqref{eq:local_cond_bare} also has a twofold dependence on $B$, suggesting an approximate factorization into valence and sea sectors again.
Specifically, we define the bare valence and sea condensates by

\begin{align}
\ave{\bar{\psi}\psi_f(x)}_{B,T}^{\mathrm{val}} &= \frac{T}{L_yL_z}\frac{1}{\Z(0)}\int \mathcal{D}U\hspace{0.1cm} e^{-S_g}\prod_{f'}\left[\det(\slashed{D}(U,0)+m_{f'})\right]^{1/4}\widehat\Tr\left[(\slashed{D}(U,q_{f}B)+m_f)^{-1}\right]\,,\label{eq:valence-chiral-condensate}\\
\ave{\bar{\psi}\psi_f(x)}_{B,T}^{\mathrm{sea}} &= \frac{T}{L_yL_z}\frac{1}{\Z(B)}\int \mathcal{D}U\hspace{0.1cm} e^{-S_g}\prod_{f'}\left[\det(\slashed{D}(U,q_{f'}B)+m_{f'})\right]^{1/4}\widehat\Tr\left[(\slashed{D}(U,0)+m_f)^{-1}\right]\,,\label{eq:sea-chiral-condensate}
\end{align}
where the magnetic field was switched off in the determinant or in the trace, respectively. Using these observables, we define the relative change of the chiral condensate with respect to the zero-field one due to the valence and the sea contributions
\begin{equation}
r^{\rm val}_f(x) = \frac{\ave{\bar{\psi}\psi_f(x)}^{\rm val}_{B,T}}{\ave{\bar{\psi}\psi_f(x)}_{0,T}} - 1 \hspace{1cm}
r^{\rm sea}_f(x) = \frac{\ave{\bar{\psi}\psi_f(x)}^{\rm sea}_{B,T}}{\ave{\bar{\psi}\psi_f(x)}_{0,T}} - 1
\label{eq:valence_sea}
\end{equation}
We can also define the light-quark average of these observables as $r^{\rm val} = (r^{\rm val}_u + r^{\rm val}_d)/2$ and $r^{\rm sea} = (r^{\rm sea}_u + r^{\rm sea}_d)/2$. For weak fields, the full condensate can be shown to be given by the sum of the sea and valence terms~\cite{d2011chiral}. For larger values of $B$, this factorization is not exact anymore, but still insightful. 

As we will see below, the impact of the two effects is quite different on the condensate: the valence effect acts in a predominantly local manner, while the sea effect is more global. Still, $r^{\rm sea}(x)$ can depend on $x$ if the gluon configurations that are preferred for $B(x)$ are themselves inhomogeneous. Exactly this is measured by the local Polyakov loop, which only depends on $B$ via the sea effect.

\section{Simulation setup}\label{sec:simulation_setup}

We used $N_f = 2+1$ flavors of rooted, stout smeared staggered quarks with physical masses and the tree-level improved Symanzik gauge action~\cite{Borsanyi:2010cj}.
The electric charges of the quarks are set to $q_u=2e/3$, $q_d=q_s=-e/3$, 
where $e>0$ is the elementary electric charge 
and the quark masses in lattice units are tuned along the line of constant physics~\cite{Borsanyi:2010cj}.

We generated gauge configurations with this action on four lattice sizes: $16^3\times6$, $24^3\times8$, $28^3\times10$ and $36^3\times12$ to approach the continuum limit. We varied the coupling $\beta$ to cover temperatures from 113 MeV to 176 MeV, in order to explore the behavior of QCD matter around the crossover temperature $T_c\approx155$ MeV. Similarly, we vary the flux quantum $N_b$, assuming a background field given by Eq.~\eqref{eq:inv_cosh_profile}, to cover values of $\sqrt{eB}$ from 0 GeV to 1.2 GeV, which is around the phenomenologically relevant strengths.
We fixed the width of the profile in physical units to $\epsilon\approx0.6$ fm, which produced an appreciable inhomogeneity in the field on the lattice and is consistent with the characteristic width of the field that was found in heavy-ion collision simulations~\cite{deng2012event}.

We remark that our lattices correspond to a fixed physical volume $V = L_xL_yL_z$. In contrast to standard thermodynamics studies, where the infinite volume limit of homogeneous systems is of interest, here we do not wish to approach that limit but are interested in the local spatial dependence of observables.

For each set of parameters, we measured the local quark condensate and the local Polyakov loop.
To calculate the condensate, we need to compute the inverse of the massive Dirac operator $\mathcal{M} = \slashed{D}(U,q_fB)+m_f$. We use the standard technique of noisy estimators, i.e.\ $N$ complex vectors whose components $\chi_k(x,y,z,t,c)$ are randomly sampled from a normal distribution, such that
\begin{equation}
\frac{1}{N}\sum_{k=1}^{N}\chi_k(x,y,z,t,c)\,\chi^*_k(x',y',z',t',c')
\delta_{xx'}\delta_{yy'}\delta_{zz'}\delta_{tt'}\delta_{cc'} + \mathcal{O}(1/\sqrt{N})\,,
\end{equation}
where the correction terms $\mathcal{O}(1/\sqrt{N})$ vanish in the limit $N\to\infty$. Here, $x$, $y$, $z$ and $t$ are coordinates and $c$ a color index. The projected trace, appearing in~\eqref{eq:local_cond_bare}, can then be approximated using these random vectors as
\begin{equation}
 \widehat\Tr(\mathcal{M}^{-1})(x) \approx \frac{1}{N}\sum_{k=1}^{N}\sum_{y,z,t,c}\chi_k^*(x,y,z,t,c) \left[\mathcal{M}^{-1}\chi_k\right](x,y,z,t,c)\,.
\end{equation}
We found that $N=40$ provides a reasonably good estimate for the local condensates. Finally, the lattice data are extrapolated to the continuum limit via a combined fit that is explained in detail in App.~\ref{app:data_continuum_limit}.
\section{Results}\label{sec:results}
Due to the inhomogeneity of $B(x)$, different parts of the system are influenced by magnetic catalysis and inverse catalysis differently, depending on the local value of the magnetic field. As a consequence, non-trivial features appear in the observables -- this is what we explore next. 
\subsection{Local magnetic catalysis}
\begin{figure}[b]
    \centering
    \includegraphics[width=0.43\linewidth]{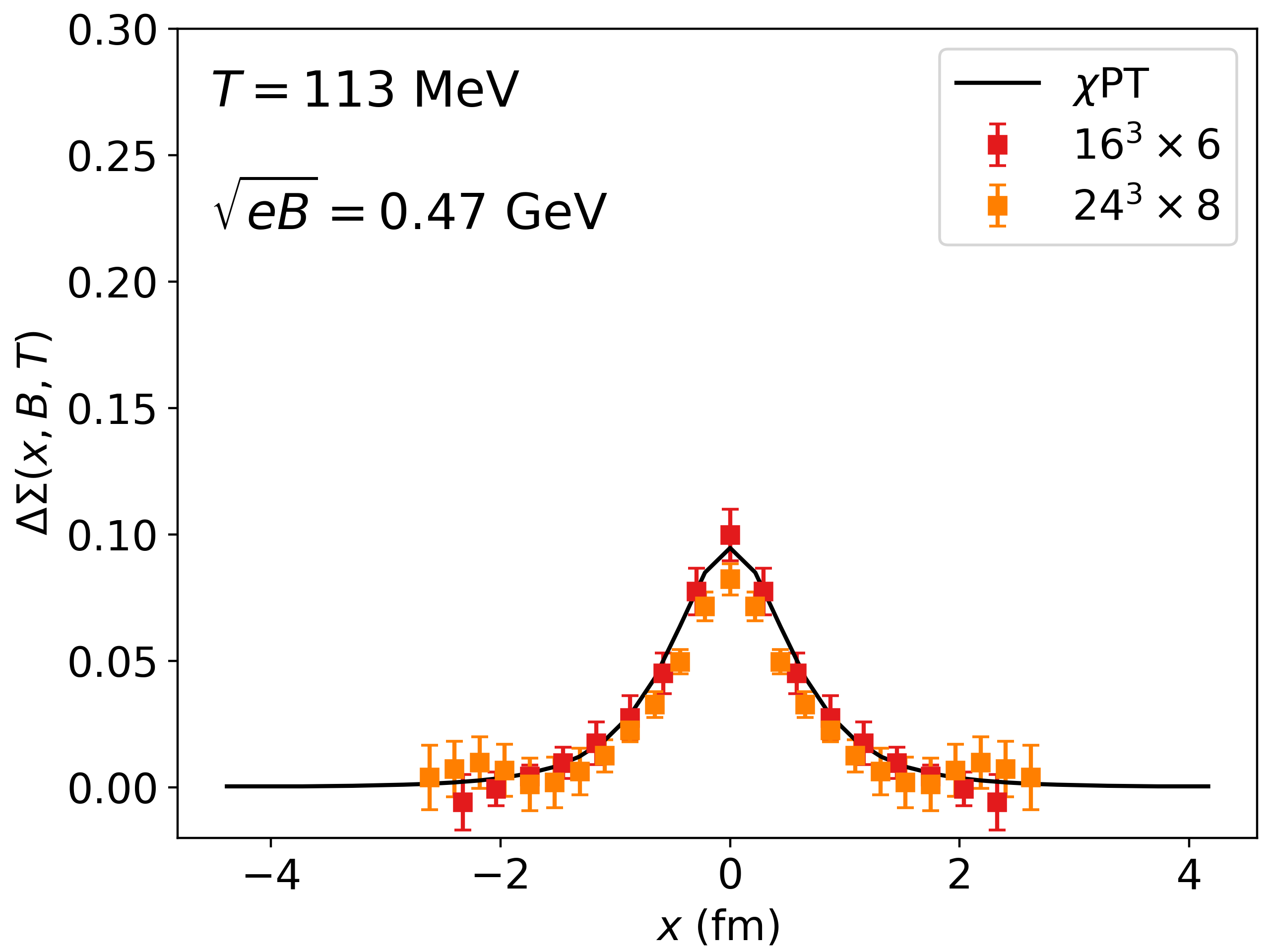}
    \quad
    \includegraphics[width=0.43\linewidth]{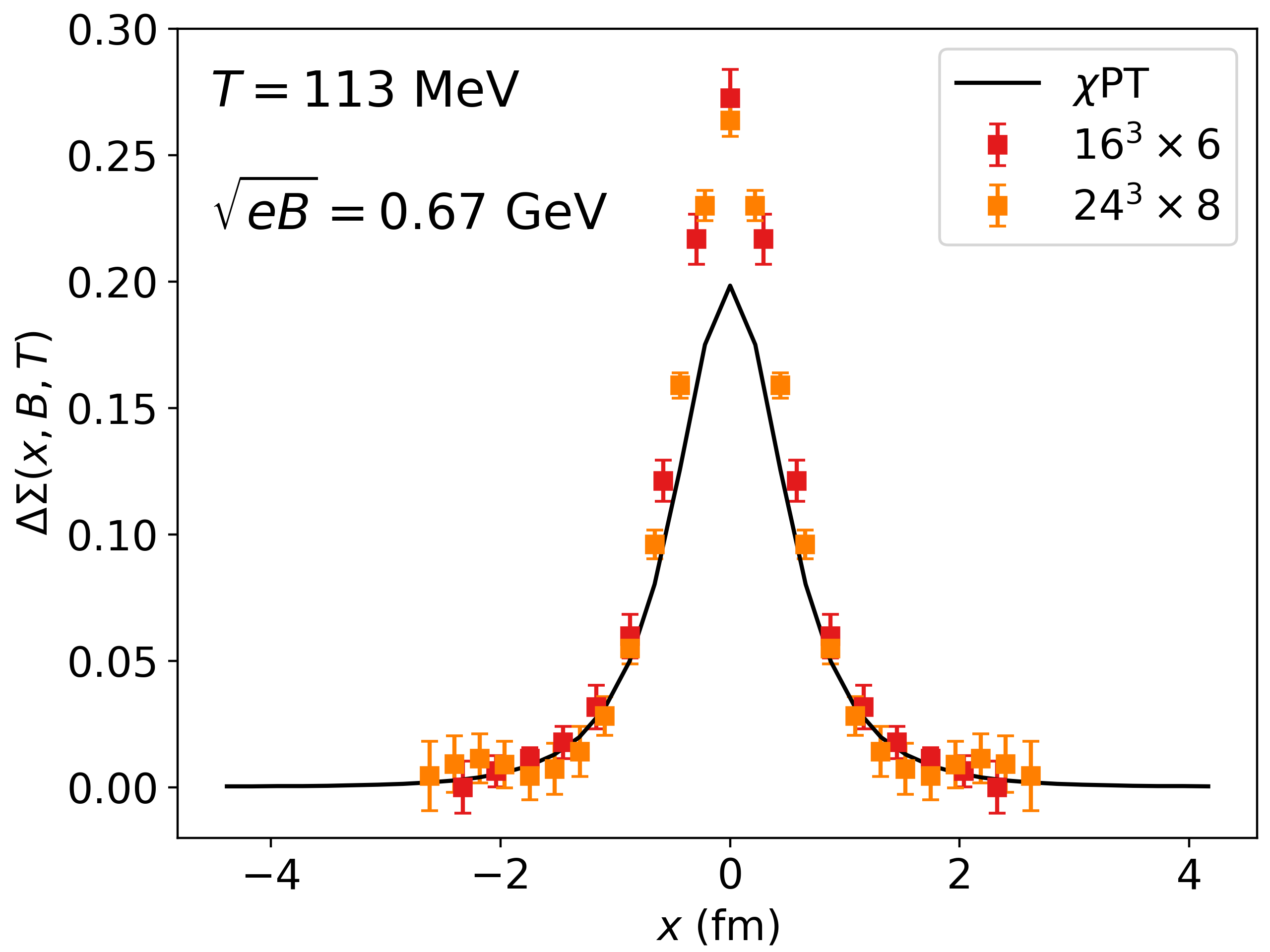}
    \caption{Modification of the renormalized chiral condensate due to the magnetic field for $16^3\times6$ and $24^3\times8$ lattices compared to $x$-dependent $\chi$PT at $T = 113$ MeV for $\sqrt{eB} = 0.47$ GeV (left) and $\sqrt{eB} = 0.67$ GeV (right).
    \label{fig:chiPT_compare_lattice}}
\end{figure}
We first discuss the $B$- and $T$-dependence of the light quark condensate in our inhomogeneous setup at low temperature.
In this region, the system is dominated by pions and its response to $B(x)$ can be described by chiral perturbation theory ($\chi$PT). We work out the details of the leading-order theory in App.~\ref{app:Diracfree} and App.~\ref{app:KGfree}.

In Fig.~\ref{fig:chiPT_compare_lattice}, we plot the modification of the chiral condensate defined in Eq.~\eqref{eq:chiral_cond_shift}, together with the $\chi$PT prediction at $T=113 \textmd{ MeV}$. Notice the local enhancement of $\Sigma(x)$ due to $B(x)$ -- a behavior that we dub {\it local magnetic catalysis}. This local increase in the condensate is described perfectly by $\chi$PT at low $B$ (left plot), while the full QCD results start to deviate from its prediction as the magnetic field increases. These deviations are more pronounced at $x=0$, where $B$ is the strongest. Altogether, we thus find that the inhomogeneous variant of $\chi$PT gives a good description of the low-temperature and low-magnetic field region, similarly to the conclusions drawn for the homogeneous case~\cite{Bali:2012zg}. 

To quantify how much the effect on the condensate follows the local profile of the magnetic field, we define an auxiliary observable, constructed solely from the data for the chiral condensate obtained for homogeneous magnetic fields~\cite{Bali:2012zg}, as
\begin{equation}
\Sigma_f^{\rm pt}(B(x),T) = \frac{2m_{f}}{m_{\pi}^2F^2}\qty[\ave{\bar{\psi}\psi_f}_{B=B(x),T} - \ave{\bar{\psi}\psi_f}_{B=T=0}]+1\,,
\label{eq:Sigma_pointwise}
\end{equation}
where $B(x)$ is our $1/\cosh^2(x)$ profile.
We also implicitly define the light-quark average of this observable. The so defined quantity -- which we name the point-wise chiral condensate -- mimics a completely local response, where the observable at point $x$ only depends on the local strength of $B$ at that point.

\begin{figure}[t]
    \centering
    \includegraphics[width=0.43\linewidth]{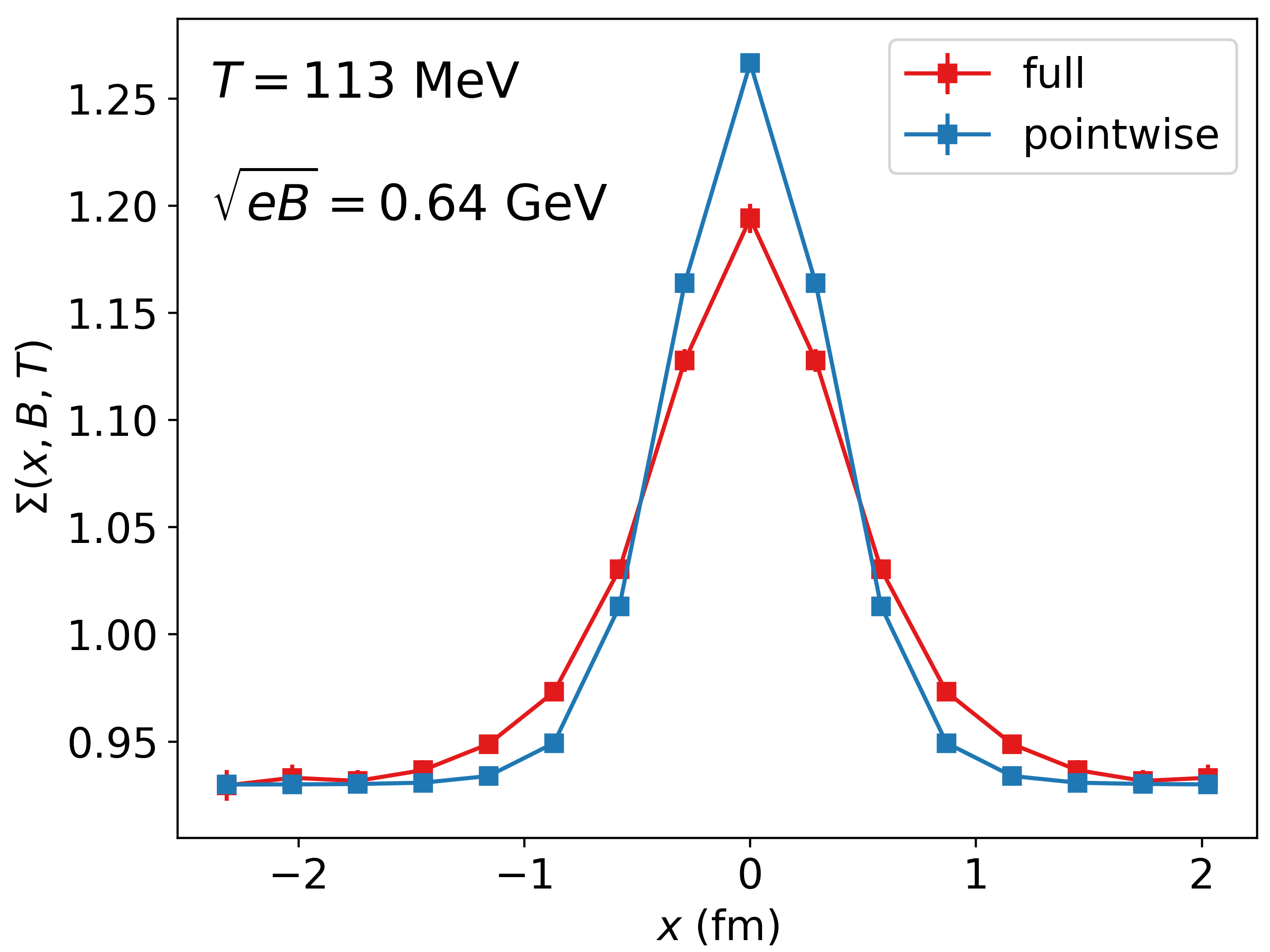}\quad
    \includegraphics[width=0.43\linewidth]{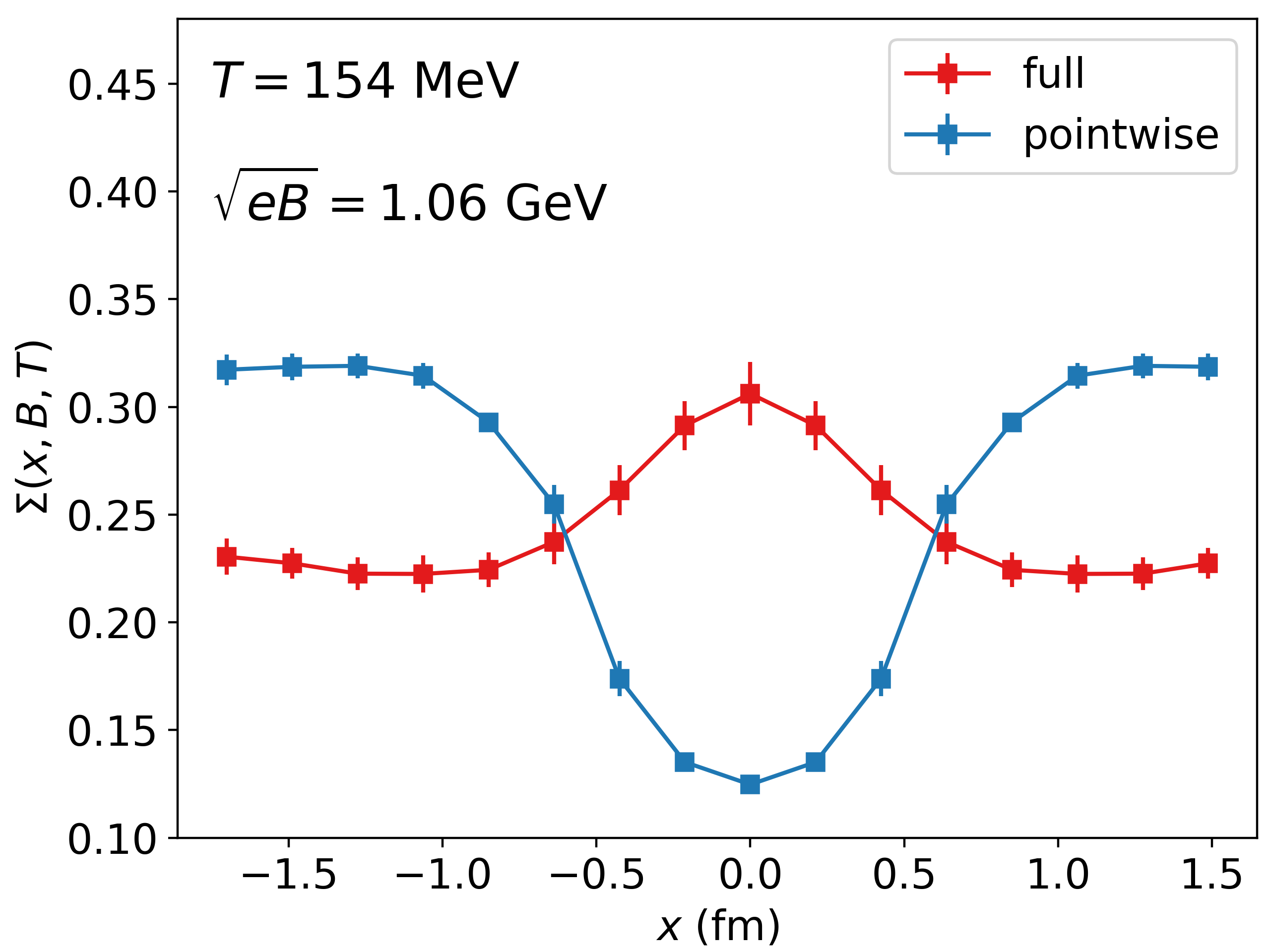}
    \caption{Point-wise condensate compared to the full one for low $T$ (left plot) and $T \approx T_c$ (right plot) on a $16^3\times6$ lattice. The peaks of the point-wise and the full condensates differ by $\sim 6\%$ at $T = 113$ MeV and by $\sim 60\%$ at $T = 155$ MeV.}
    \label{fig:ptwise_vs_full}
\end{figure}

In Fig.~\ref{fig:ptwise_vs_full}, we compare the point-wise observable with the full condensate~\eqref{eq:Sigmafdef1} at different temperatures.
At low $T$ (left plot), the difference between the peaks of the condensates is only about $6\%$, which means that the full condensate feels the magnetic field approximately in a local manner. However, in the transition region (right plot), the point-wise condensate would lead to the wrong behavior, as it decreases abruptly at $x = 0$ compared to the full one. The difference in the peaks is $\sim60\%$ and it indicates that the global effects of $B$ are much more manifest around the crossover.

\subsection{Local inverse magnetic catalysis}

\begin{figure}[t]
    \centering
    \includegraphics[width=0.43\linewidth]{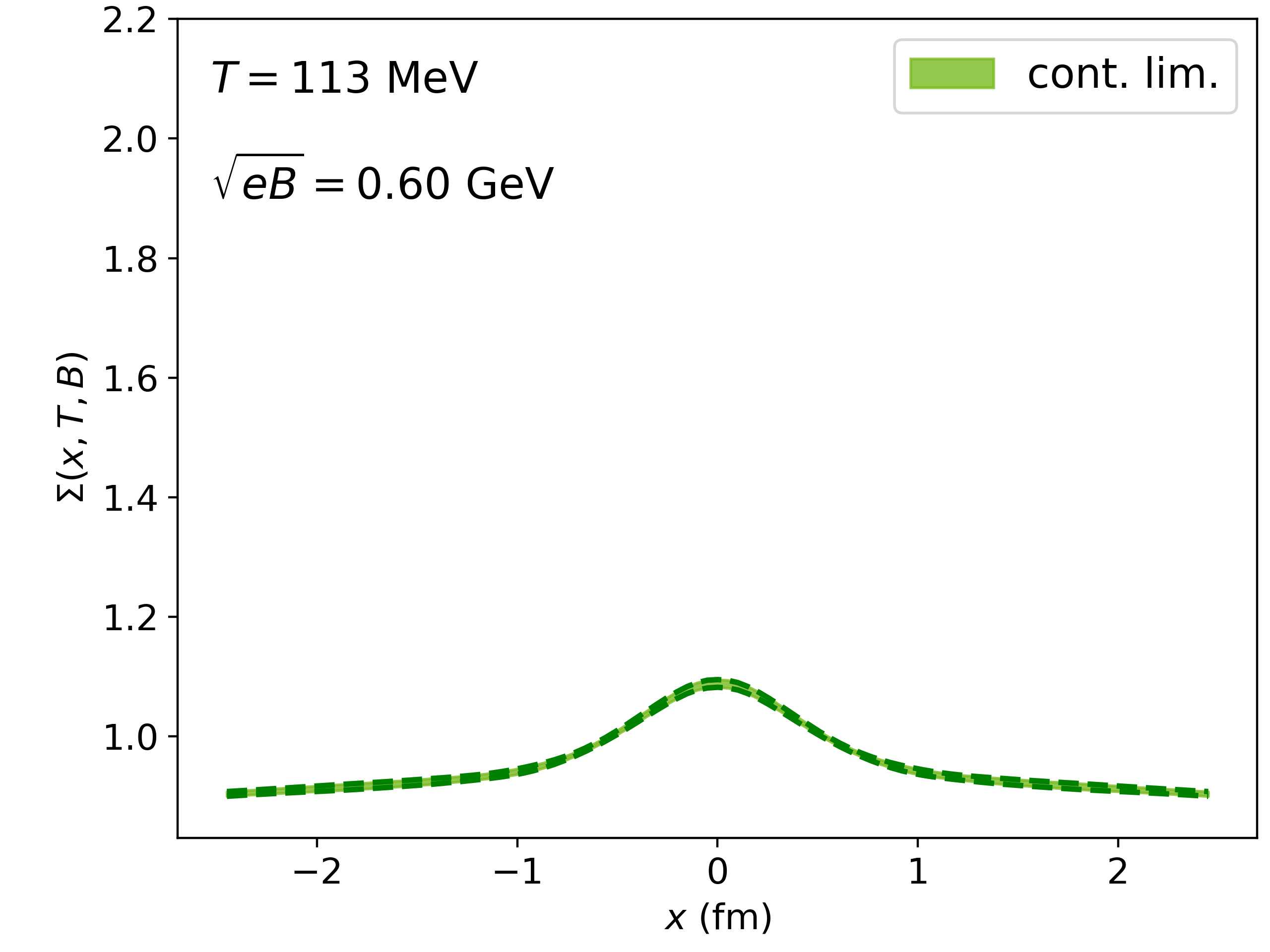}\quad
    \includegraphics[width=0.43\linewidth]{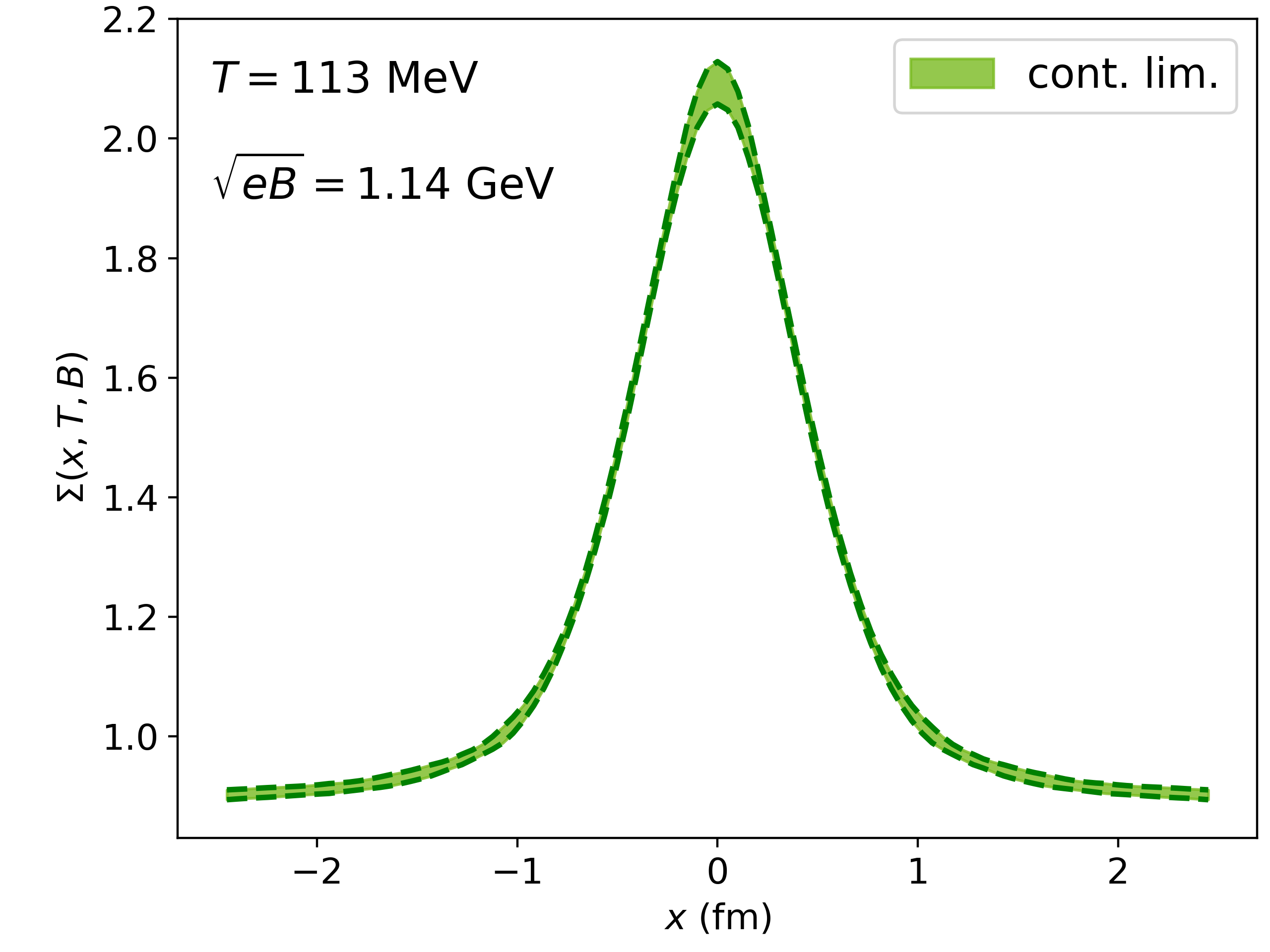}
    \includegraphics[width=0.43\linewidth]{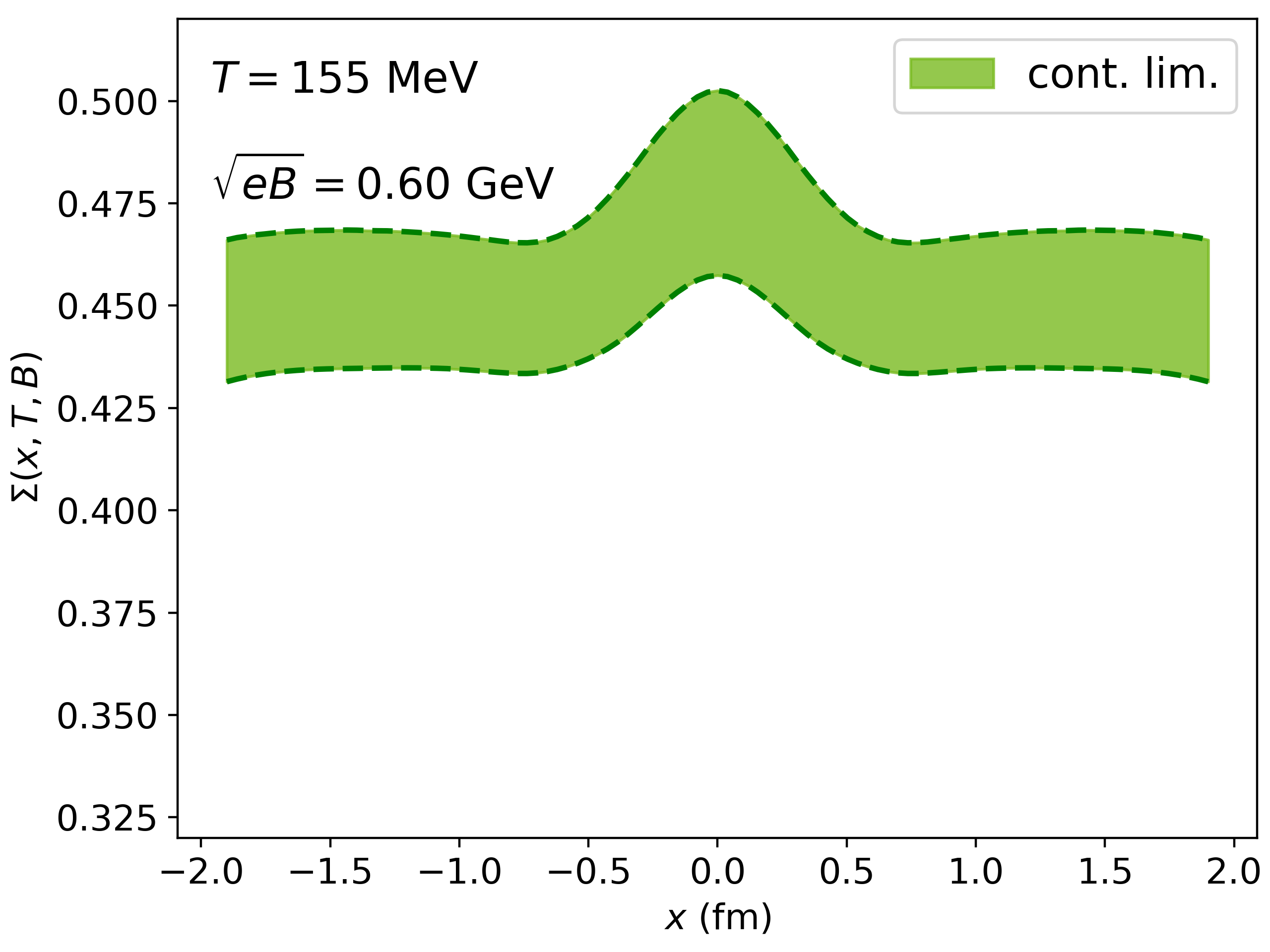}\quad
    \includegraphics[width=0.43\linewidth]{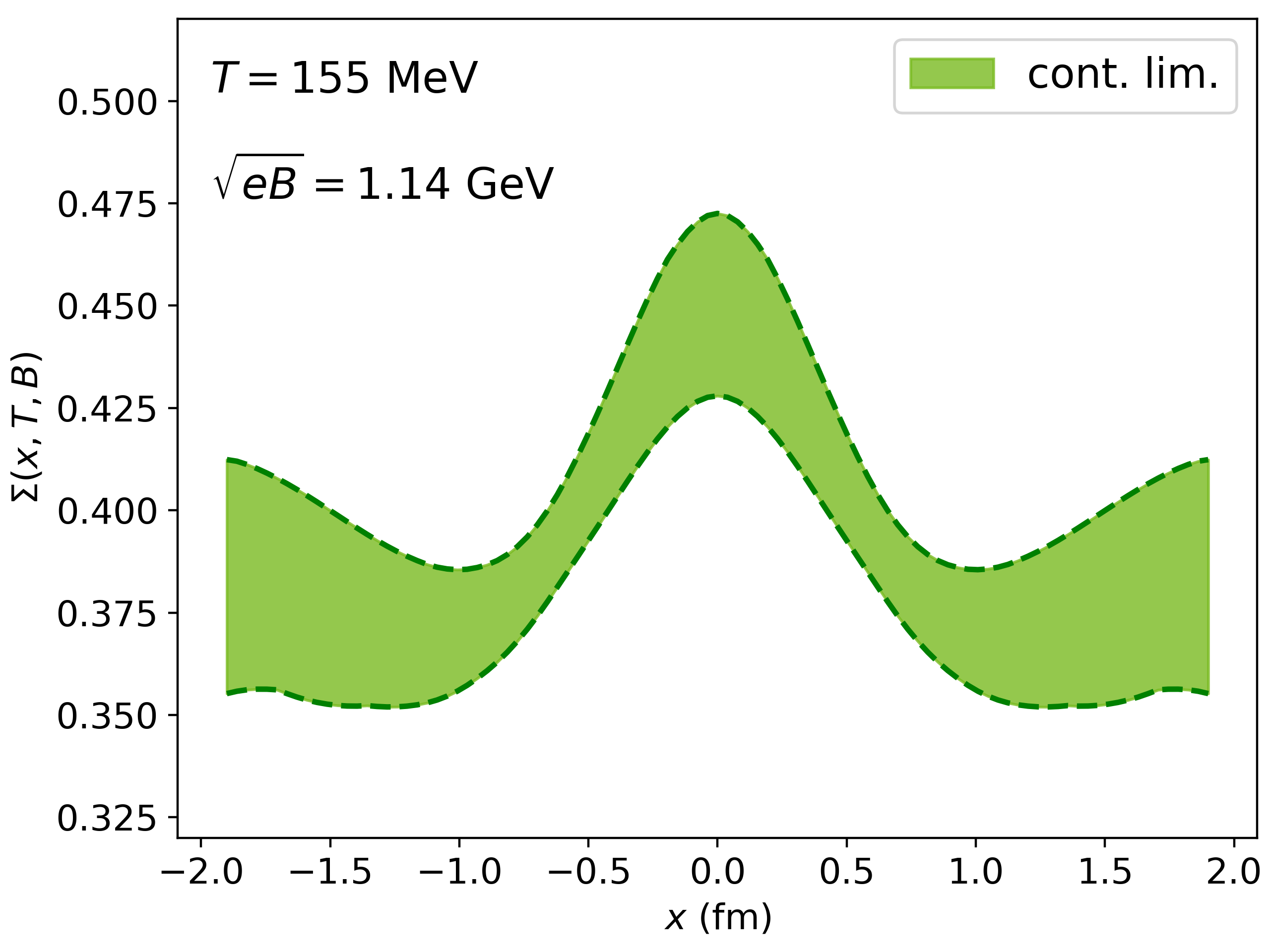}
    \caption{Continuum extrapolated results for the renormalized quark condensate as a function of $x$ for different magnetic fields and temperatures.
    \label{fig:contlim_chiral_condensate}}
\end{figure}

\begin{figure}[b]
    \centering
    \includegraphics[width=0.43\linewidth]{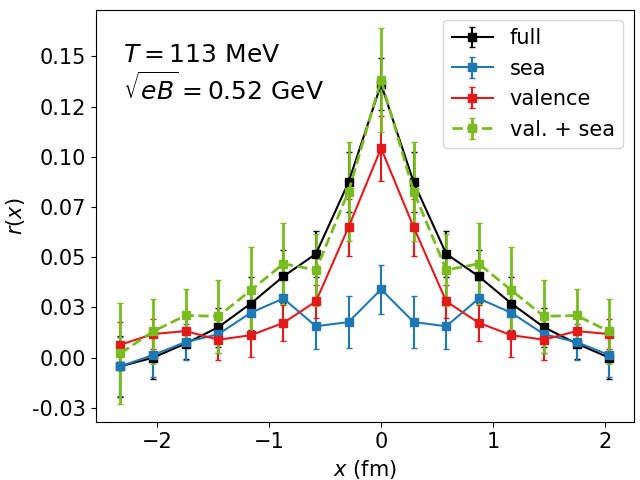}\quad
    \includegraphics[width=0.43\linewidth]{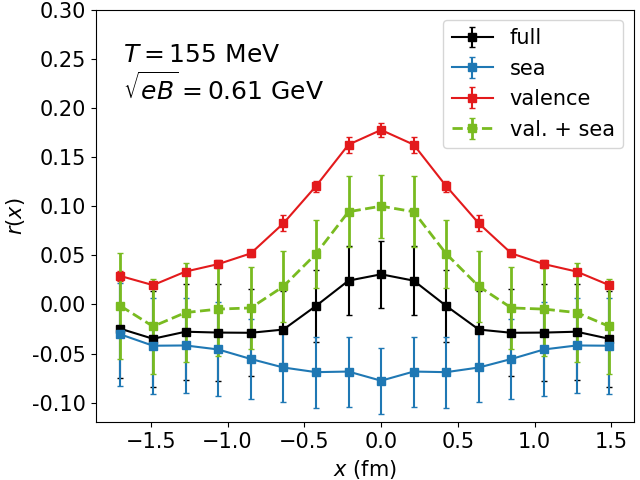}
    \caption{Sea and valence contributions to the chiral condensate compared to the full one for $T = 113$ MeV and $T = 155$ MeV.
    \label{fig:sea_valence}}
\end{figure}

Next, we discuss our full results for $\Sigma(x)$ in more detail.

A continuum extrapolation based on four different lattice spacings is 
carried out (see details in App.~\ref{app:data_continuum_limit}) and here we work directly with the results obtained for $a\to0$. This is shown in Fig.~\ref{fig:contlim_chiral_condensate},which constitutes one of our main results.

Tabulated data for the continuum extrapolation is available as part of this publication and will serve as a non-trivial benchmark for low-energy models and effective theories of QCD.
We note that there are substantial correlations between $\Sigma(x)$ at different coordinates that would make the error of, for example $\Sigma(x)-\Sigma(0)$ smaller than that of $\Sigma(x)$ itself. Nevertheless, we chose to work with the unsubtracted observable to enable a complete comparison to other approaches.

In Fig.~\ref{fig:contlim_chiral_condensate} we consider two temperatures: $T= 113\textmd{ MeV}$ and $T= 155\textmd{ MeV}$. Based on results obtained for homogeneous fields~\cite{Bruckmann:2013oba}, the valence effect is expected to be dominant at the lower temperature, while a competition between the valence and sea effects should take place at the higher temperature.
In the upper plots ($T=113$ MeV), we again see the peak of the local condensate, following closely the $1/\!\cosh^2$ field profile -- a manifestation of the fact that the valence effect impacts locally on the condensate. In turn, the behavior of $\Sigma(x)$ at $T=155$ MeV is more complicated. First, we observe an overall decrease of the condensate for increasing $B$. Second, $\Sigma(x)$ is also deformed and we observe the formation of dips in the tails of the curves, i.e., localized regions where $\Sigma(x)$ is smaller even compared to the edges of the lattice, where the magnetic field practically vanishes. This nontrivial behavior is due to the interplay of the sea effect (reducing the condensate globally) and the valence effect (increasing it locally) that we will explain in more detail below.

The appearance of the dips, and the nontrivial behavior of $\Sigma(x)$ is a new type of indication of the inverse magnetic catalysis phenomenon.

There are two avenues to understand this behavior better: via a separate analysis of the valence and sea contributions to $\Sigma(x)$, and via an investigation of the Polyakov loop $P_R(x)$.
First we discuss the contributions from the sea and the valence effects, as defined in Eq.~\eqref{eq:valence_sea}. 
These are plotted in Fig.~\ref{fig:sea_valence} and compared to the full condensate.
At low temperature (left plot), the dominant contribution is clearly given by the valence part, whereas the sea contribution is only slightly above zero. On the other hand, at $T \approx T_c$ the situation changes and the sea contribution becomes not only relevant but also negative, therefore acting to decrease the condensate. For even higher $B$, the negative contribution of the sea effect tends to dominate over the valence effect, although in this regime the additive property of the two effects no longer holds. 

Notice that the sea contribution in both cases tends to be more broadly distributed, whereas the valence one is more sharply peaked. That difference is essential for the appearance of the dips in the chiral condensate and can be explained via the Polyakov loop, which we discuss in the next section.

\subsection{Local Polyakov loop}

\begin{figure}[b]
    \centering
    \includegraphics[width=0.43\linewidth]{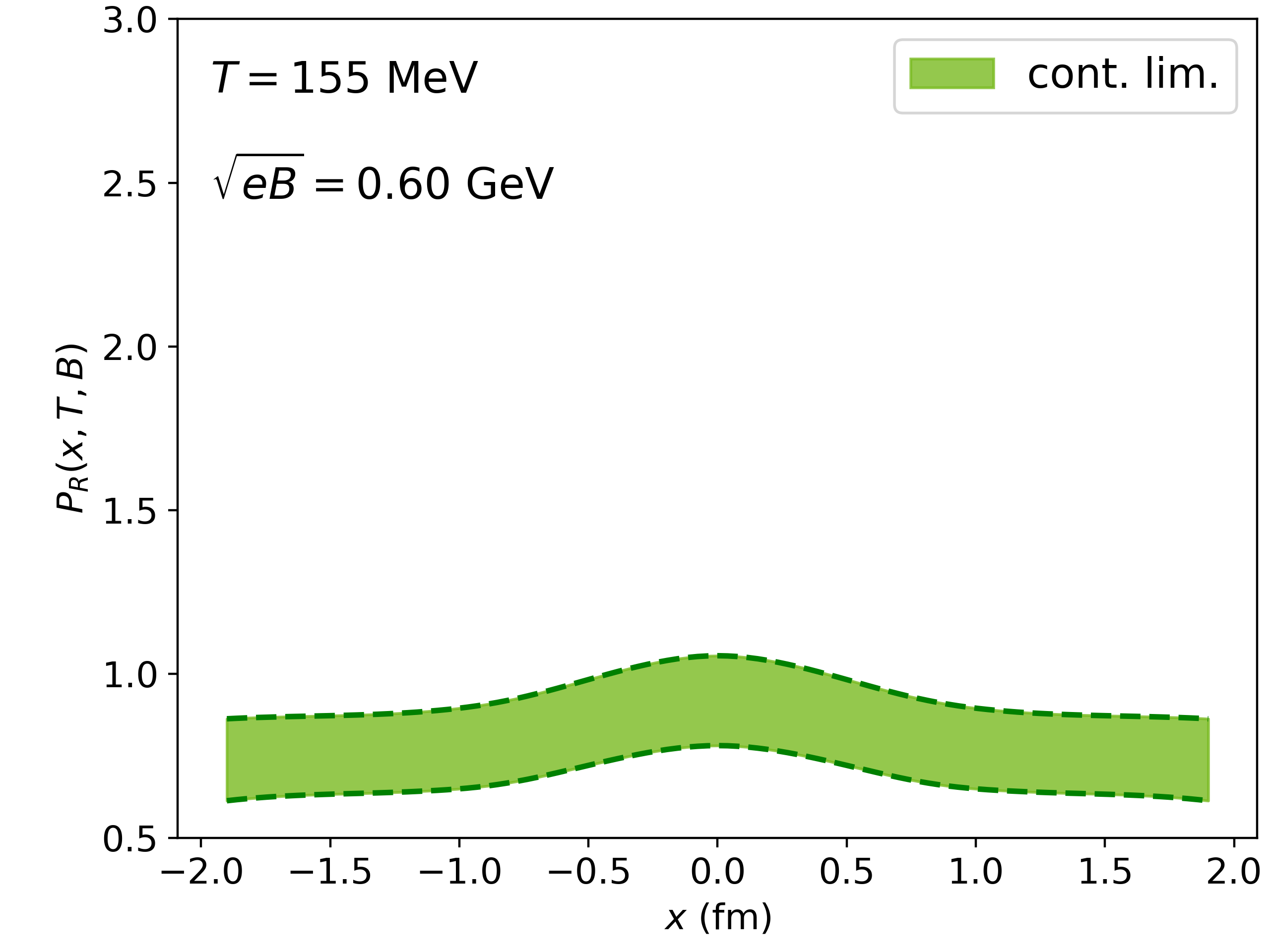}
    \quad
    \includegraphics[width=0.43\linewidth]{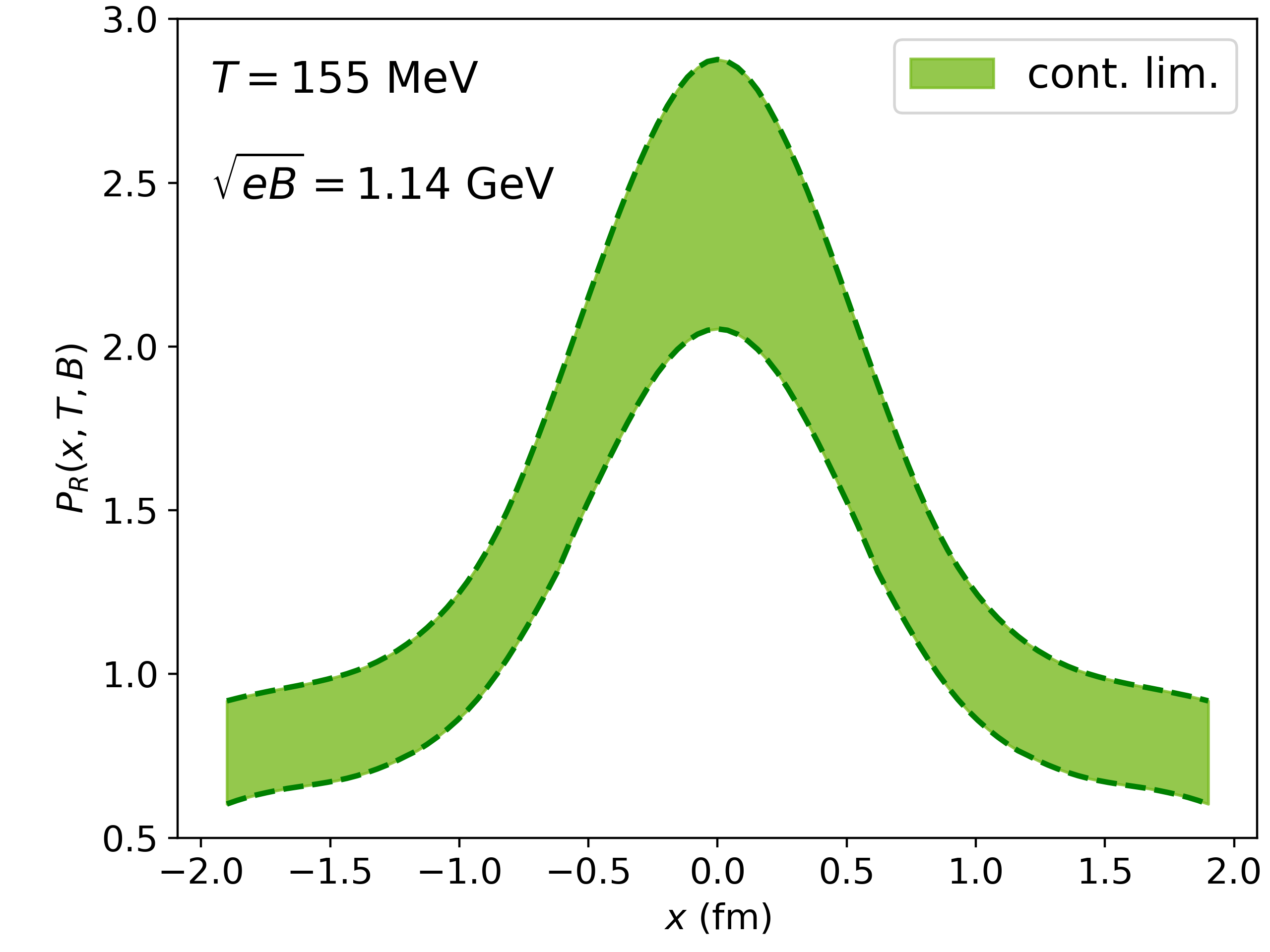}
    \caption{Continuum extrapolated renormalized Polyakov loop as a function of $x$ at $T = 155$ MeV for $\sqrt{eB} = 1$ GeV (left) and $\sqrt{eB} = 1.15$ GeV (right).
    \label{fig:ploop_x}}
\end{figure}

The sea effect is known to be rooted in the behavior of the Polyakov loop; therefore we turn to this observable next. Its continuum extrapolation is also  discussed in App.~\ref{app:data_continuum_limit}.
In Fig.~\ref{fig:ploop_x}, we show the continuum extrapolated, renormalized local Polyakov loop as a function of $x$ for two different magnetic fields at the temperature ($T=155$ MeV), where we found the dips to develop in the light quark condensate. We first notice that the Polyakov loop is also enhanced where the magnetic field is stronger. As we discussed above, the Polyakov loop is a purely gluonic quantity that does not couple to the magnetic field directly, and is only affected by the sea effect through the fermion determinant. Since the latter depends on the magnetic field at all lattice points, this implies that $B$ has a smeared impact on $P$. Therefore, the Polyakov loop feels the magnetic field in a broader region of the lattice compared to the quark condensate. This is manifested by the broader peaks in $P_R(x)$ in the figure.

\begin{figure}[t]
    \centering
    \includegraphics[width=0.43\linewidth]{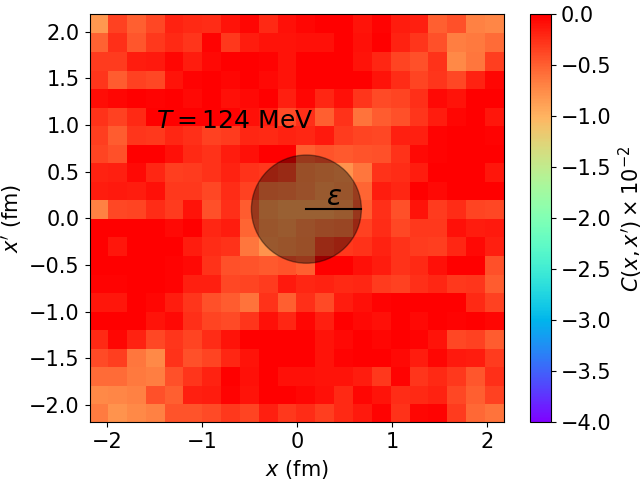} \quad
    \includegraphics[width=0.43\linewidth]{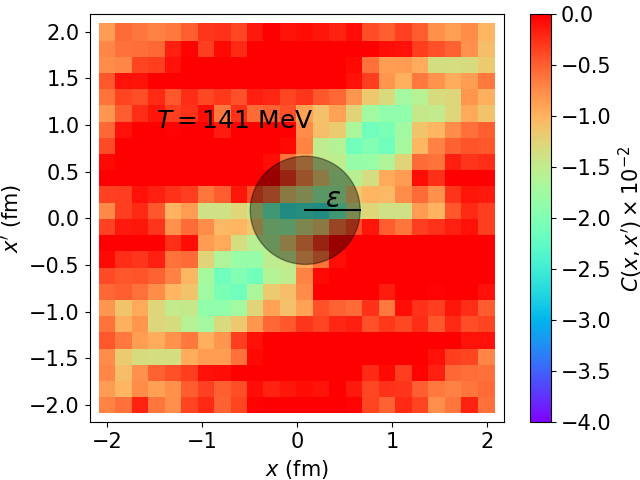}
    \includegraphics[width=0.43\linewidth]{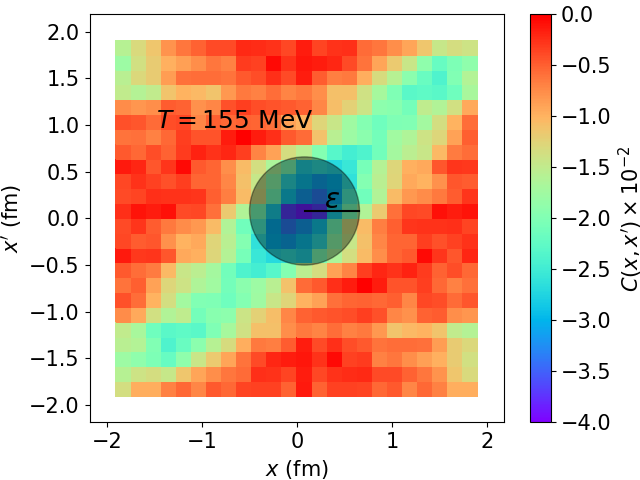} \quad
    \includegraphics[width=0.43\linewidth]{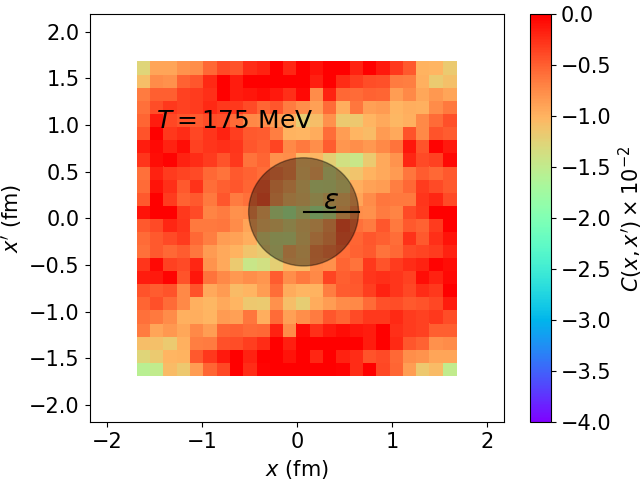}
    \caption{Heat plot of the correlation $C(x,x')$ between the bare quark condensate (at point $x$) and the bare Polyakov loop (at point $x'$) for temperatures $T<T_c$ (upper left panel), $T \approx T_c$ (upper right and lower left panels) and $T>T_c$ (lower right panel) on a $24^3\times8$ lattice. The rough size $\epsilon$ of the magnetic field is indicated by the gray circle. The scale of the plot is kept fixed in physical units, implying that the heat map area slightly shrinks as $a$ reduces towards higher temperatures.
    \label{fig:cp_correlation_T123}}
\end{figure}

What is the impact of such a Polyakov loop profile on the condensate?
In the homogeneous case, it is known that $P$ and $\bar\psi\psi$ anticorrelate with each other in the transition region~\cite{Bruckmann:2013oba}. 
To answer this question for our inhomogeneous setup, we need to quantify the extent of {\it local} correlations between the two observables. To this end, we define the 
correlator of 
the bare quark condensate and the bare Polyakov loop
\begin{equation}
C(x,x^{\prime}) = \frac{1}{m_{\pi}^3}\qty[\ave{\bar{\psi}\psi(x)P(x^{\prime})}-\ave{\bar{\psi}\psi(x)}\ave{P(x^{\prime})}]\,,
\end{equation}
where we employed a convenient normalization to make $C$ dimensionless.\footnote{Note that due to the breaking of translational invariance by the magnetic field, $C(x,x')$ depends on both coordinates and not just on their difference. Note also that parity symmetry ensures $C(-x,-x')=C(x,x')$, but $C(x,x')\neq C(x',x)$ due to the difference of the two operators at the two positions.}
In Fig.~\ref{fig:cp_correlation_T123}, we show a heat plot of $C(x,x^{\prime})$ for different temperatures at approximately the same magnetic field. 

We can see that there are no sizeable correlations at low or high temperatures. In contrast, there is a strong local anticorrelation between the Polyakov loop and the condensate in the transition region. This anticorrelation is most prominent along the diagonal of the figure, i.e.\ when both quantities are taken at the same coordinate, but the negative correlation persists even away from the diagonal. Thus, the suppressing effect of the Polyakov loop can impact the condensate from a distance of about $0.5\textmd{ fm}$.

\begin{figure}[!h]
    \centering
    \includegraphics[width=0.49\linewidth]{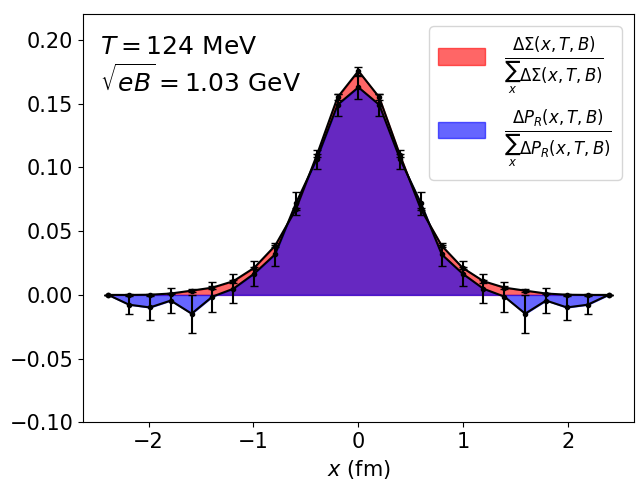}
    \includegraphics[width=0.49\linewidth]{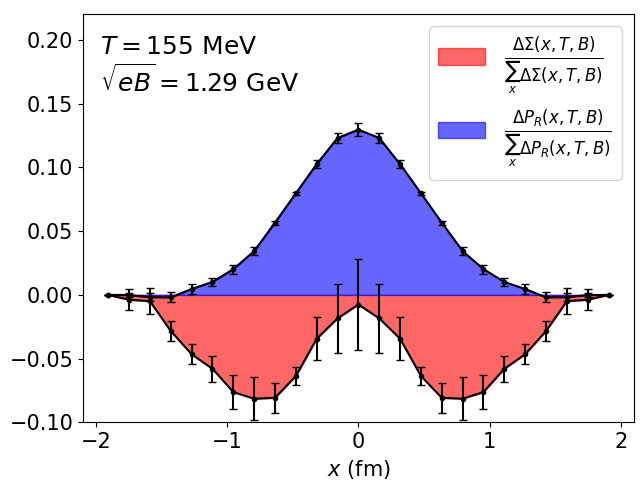}
    \includegraphics[width=0.49\linewidth]{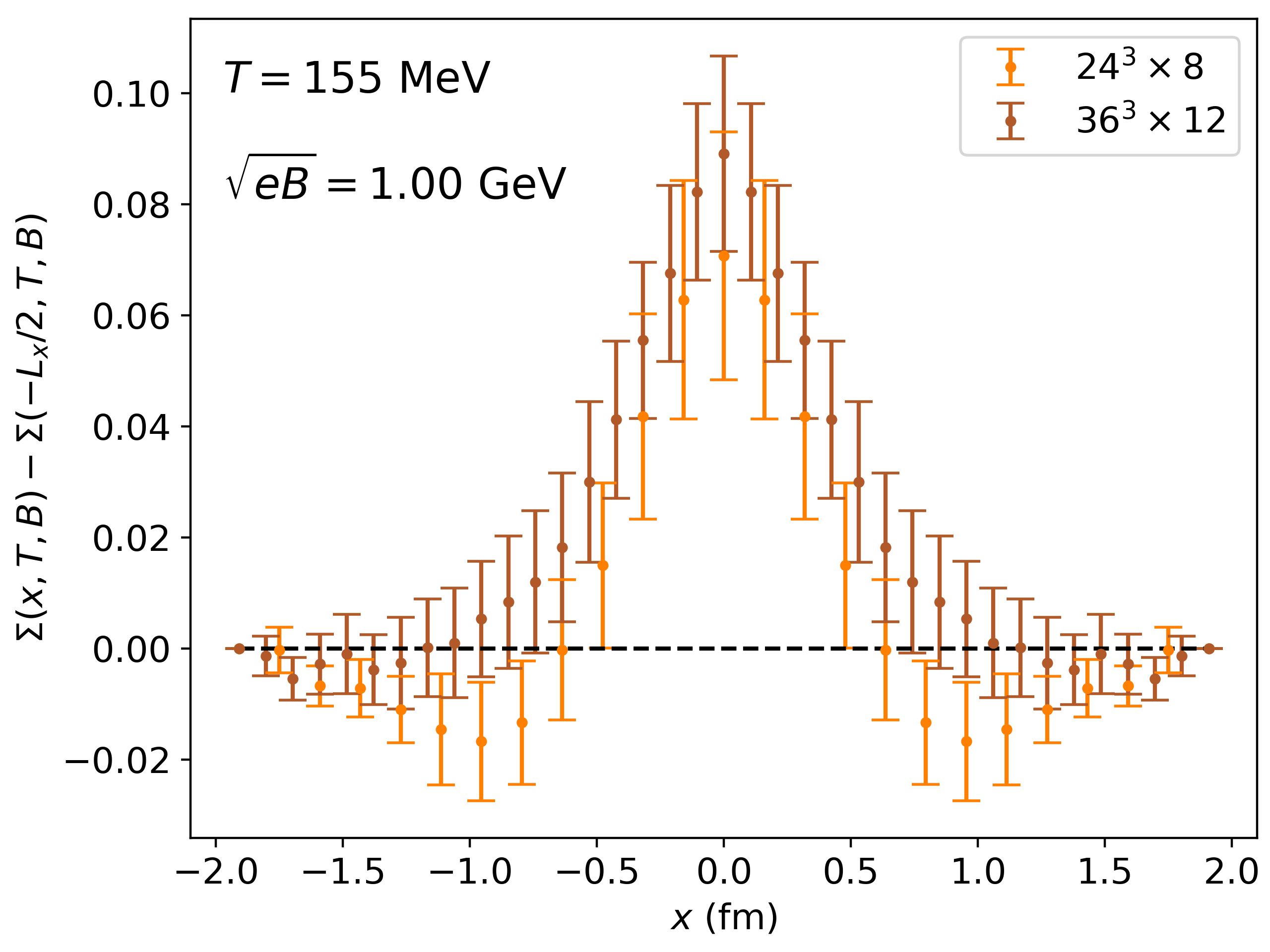}
    \includegraphics[width=0.49\linewidth]{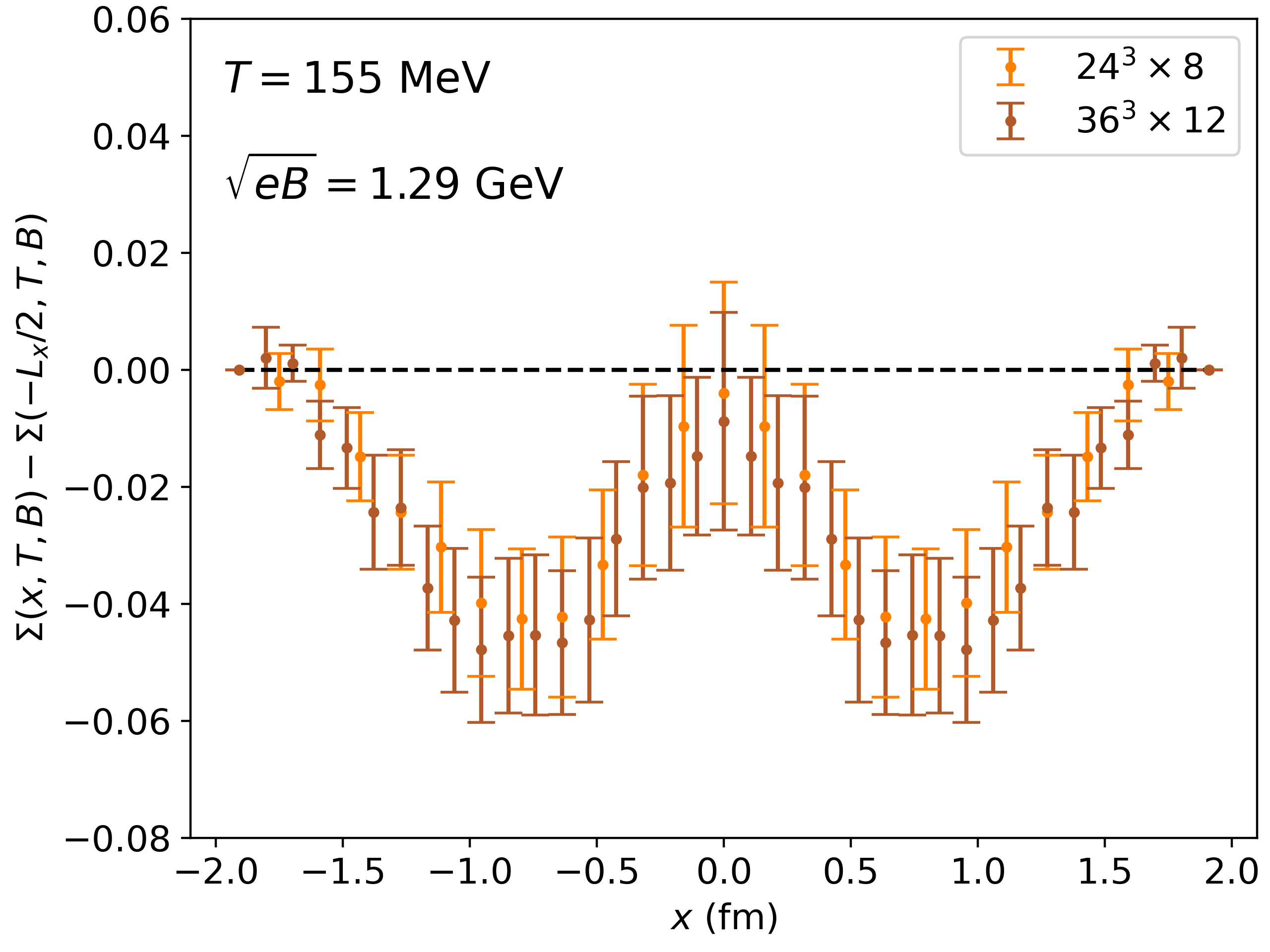}
    \caption{Top panels: Local profiles of the quark condensate (red) and the Polyakov loop (blue) on a $28^3 \times 10$ lattice at $T=124 \textmd{ MeV}$ (left) and at $T=155\textmd{ MeV}$ (right). For the sake of visualization, we subtracted the values of both observables at the edge, $x=\pm L_x/2$ (denoted by $\Delta$) and normalized by the area. Notice the broader peak of the Polyakov loop and the formation of the dips in the condensate in the transition region. Lower panels: renormalized chiral condensate at the point $x$ minus the one at the edge ($\pm L_x/2$) showing the formation of the dips for increasing magnetic fields on the $N_t = 8$ and $N_t = 12$ lattices.}
    \label{fig:cond_ploop_comparison}
\end{figure}

We can now put together the valence and sea effects for the condensate. The valence effect enhances $\Sigma^{\rm val}(x)$ in roughly the same shape as that of the magnetic field $B(x)$. The sea effect acts via the Polyakov loop, which feels the magnetic field in a smeared (global) way and is enhanced in a broader region. The anticorrelation with the local condensate thus suppresses $\Sigma^{\rm sea}(x)$ in a broader interval, too.
The total effect arises as a combination of both, explaining the formation of the dips in $\Sigma(x)$ in the transition region. This combined effect is illustrated again in Fig.~\ref{fig:cond_ploop_comparison}, where we show a comparison between the renormalized quantities, namely the Polyakov loop and the chiral condensate, normalized by their sum over the whole lattice for $T < T_c$ and $T \sim T_c$. To highlight the local variation of the observables, here we subtract their values at the edge of the volume $x=\pm L_x/2$.

\section{Summary and conclusions}\label{sec:conclusions}
In this paper we presented the first lattice simulations of QCD with physical quark masses in the presence of inhomogeneous external fields. In particular, we considered a localized background magnetic field $B(x)\propto \cosh^{-2}(x/\epsilon)$. Choosing a profile width $\epsilon=0.6\textmd{ fm}$, this setup resembles the conditions relevant for the early stages of off-central heavy-ion collisions. We calculated observables relevant for the thermodynamics of the system: the local light quark condensates as well as the local Polyakov loop. Our results are provided in ancillary files submitted to \verb|arxiv.org| alongside with this publication and will serve as useful benchmarks for effective theories and low-energy models of QCD, in a similar way to the results for homogeneous fields~\cite{bali2012qcd,Bali:2012zg}.
This is in particular the case because the profile of the field allows for an analytical treatment in NJL-type models, as in Ref.~\cite{cao2018chiral}.
Our findings may also provide useful input for hydrodynamic models aiming to describe off-central heavy-ion collisions. 

At low temperatures, we established the local analogon of the well-known magnetic catalysis phenomenon: the enhancement of the local condensate due to the local magnetic field. We worked out how to treat this setup within leading-order chiral perturbation theory (for details, see App.~\ref{app:KGfree}). The corresponding prediction for the condensate was found to exhibit good agreement with the lattice results for weak magnetic fields at low temperature. Moreover, we found that this effect mainly stems from the valence sector and acts in a local manner on the condensate.

In contrast, the behavior of the condensate is much more complicated in the transition region. Here, we performed a systematic study of the dependence of the local condensate and the local Polyakov loop on the magnetic field, the temperature as well as the coordinate $x$.
Our results show an enhancement of the condensate $\Sigma(x)$ similar to the shape of $B(x)$, together with a reduction in a broader interval, leading to the development of the `dips' away from the center.
This is a novel, non-trivial manifestation of the inverse magnetic catalysis phenomenon, arising from a delicate interplay between the local Polyakov loop and the local condensate via their local anticorrelations.

In summary, our results reveal that the inhomogeneous background magnetic field can realize situations where different parts of the system feature substantially different values for the relevant thermodynamic observables. In terms of the local Polyakov loop for example, one might describe the center of the system to behave more as deconfined, whereas the outer regions more as confined matter. However, unlike in the case of homogeneous fields, this picture is more complicated due to the quark condensate being also more enhanced in the center and reducing towards the edges. In other words, the extent to which deconfinement and chiral symmetry restoration occur simultaneously, is different point by point. Due to the crossover nature of the transition, a smooth dependence of all observables on $x$ is natural. An intriguing question is whether the local variations in these observables become abrupt when magnetic fields are investigated for which the QCD transition becomes first order~\cite{DElia:2021yvk}, even though these are not relevant for heavy-ion phenomenology anymore.

\acknowledgments
The authors are grateful for enlightening discussions with Matteo Giordano and Tam\'as Kov\'acs.
We acknowledge support
by the Deutsche Forschungsgemeinschaft (DFG, German
Research Foundation) through the CRC-TR 211 ``Strong-interaction matter under extreme conditions'' -- project
number 315477589. F.C. acknowledges the
support by the State of Hesse within the Research Cluster ELEMENTS (Project ID 500/10.006). 
Parts of the computations in this work were performed on the GPU
cluster at Bielefeld University and on the Goethe-HLR cluster at Goethe University. We thank the computing staff of both institutions for their support.

\appendix
\section{Prescription for the U(1) links}
\label{app:u1links}
In this work, we have used a $1/\cosh^2$ function to model the profile in heavy-on collisions. Even though the link prescription used in Sec.~\ref{sec:mag_field} was derived for that case, in this appendix we show how to construct the links for a general magnetic field of the type $\Vec{B} = B(x,y)\hat{z}$, i.e., a $z$-oriented magnetic field which depends on both the $x$ and $y$ coordinates. This procedure can also be used in future implementations of inhomogeneous magnetic fields on the lattice. In this appendix, we consider a single quark flavor with electric charge $q$ and label the coordinates so that $-L_x/2\le x<L_x/2$ and $-L_y/2\le y<L_y/2$.

In terms of the vector potential, $B$ can be written as
\begin{equation}
B = \pdv{A_y}{x}-\pdv{A_x}{y}\,.
\end{equation}
Choosing a gauge where $A_x=0$, the equation above simplifies to $B = \partial A_y/\partial x$. Therefore
\begin{equation}
A_y(x,y) = A_y(-L_x/2,y) + \int_{-L_x/2}^x dx^{\prime}B(x^{\prime},y)
\label{eq:ay_gauge_transf}
\end{equation}
Due to the second term above, $A_y(L_x/2,y) \neq A_y(-L_x/2,y)$, and the gauge field is not periodic. However, on the lattice, it is convenient to have exact periodicity. To circumvent this issue, we need to gauge-transform $\Vec{A}$ such that all components are periodic. We first observe that the second term in~\eqref{eq:ay_gauge_transf} can be regarded as a gauge transformation of the type: $A_y(L_x/2,y) = A_y(-L_x/2,y) + \partial_y f(L_x/2,y)/q$. Solving for the gauge transforming function $f$, we obtain
\begin{equation}
f(L_x/2,y) = q\int_{-L_y/2}^y d y' \int_{-L_x/2}^{L_x/2}dx^{\prime} \,B(x^{\prime},y^{\prime})\,,
\end{equation}
where the constant term $f(L_x/2,-L_y/2)$ was chosen to be zero w.l.o.g. Since the gauge transformation function can be defined point by point, we can write $f$ on the whole lattice as 
\begin{equation}
f(x,y) = \left\{\begin{array}{ll}
     q\int_{-L_y/2}^y dy' \int_{-L_x/2}^{L_x/2}dx^{\prime}\,B(x^{\prime},y^{\prime}) \quad & \text{if } x = L_x/2\,,\\
     0 & \text{otherwise} \,.
\end{array}\right.
\label{eq:g_transform_func}
\end{equation}
Eq.~\eqref{eq:g_transform_func} gives the exact function under which $A_y$ has to transform in order to become periodic. At the level of the links $u_{\mu} = e^{iaqA_{\mu}}$, the gauge transformation is defined as $\Omega(x,y) = e^{if(x,y)}$ and it acts on the last $x$-slice in the following way
\begin{align}
u_y(L_x/2,y)^{\prime} &= \Omega(L_x/2,y)\,u_y(L_x/2,y)\,\Omega(L_x/2,y+a)^{\dagger} \nonumber \\
&= \exp[if(L_x/2,y) + iaqA_y(L_x/2,y)-if(L_x/2,y+a)] \nonumber \\
&= \exp[iaqA_y(-L_x/2,y) + iaq\int_{-L_x/2}^{L_x/2} dx^{\prime}B(x^{\prime},y) - iq\int_y^{y+a}dy'\int_{-L_x/2}^{L_x/2} dx^{\prime}\,B(x^{\prime},y^{\prime})]\,.
\end{align}
Up to $\mathcal{O}(a)$, we can show that the third term in the exponential
\begin{equation}
\int_y^{y+a} dy'\int_{-L_x/2}^{L_x/2} dx^{\prime}\,B(x^{\prime},y^{\prime}) \approx a\int_{-L_x/2}^{L_x/2}dx^{\prime}B(x^{\prime},y)\,,
\end{equation}
which cancels the second one, yielding $u_y(L_x/2,y) = u_y(-L_x/2,y)$ as expected. 

\begin{figure}[!h]
    \centering
    \includegraphics[width=0.8\textwidth]{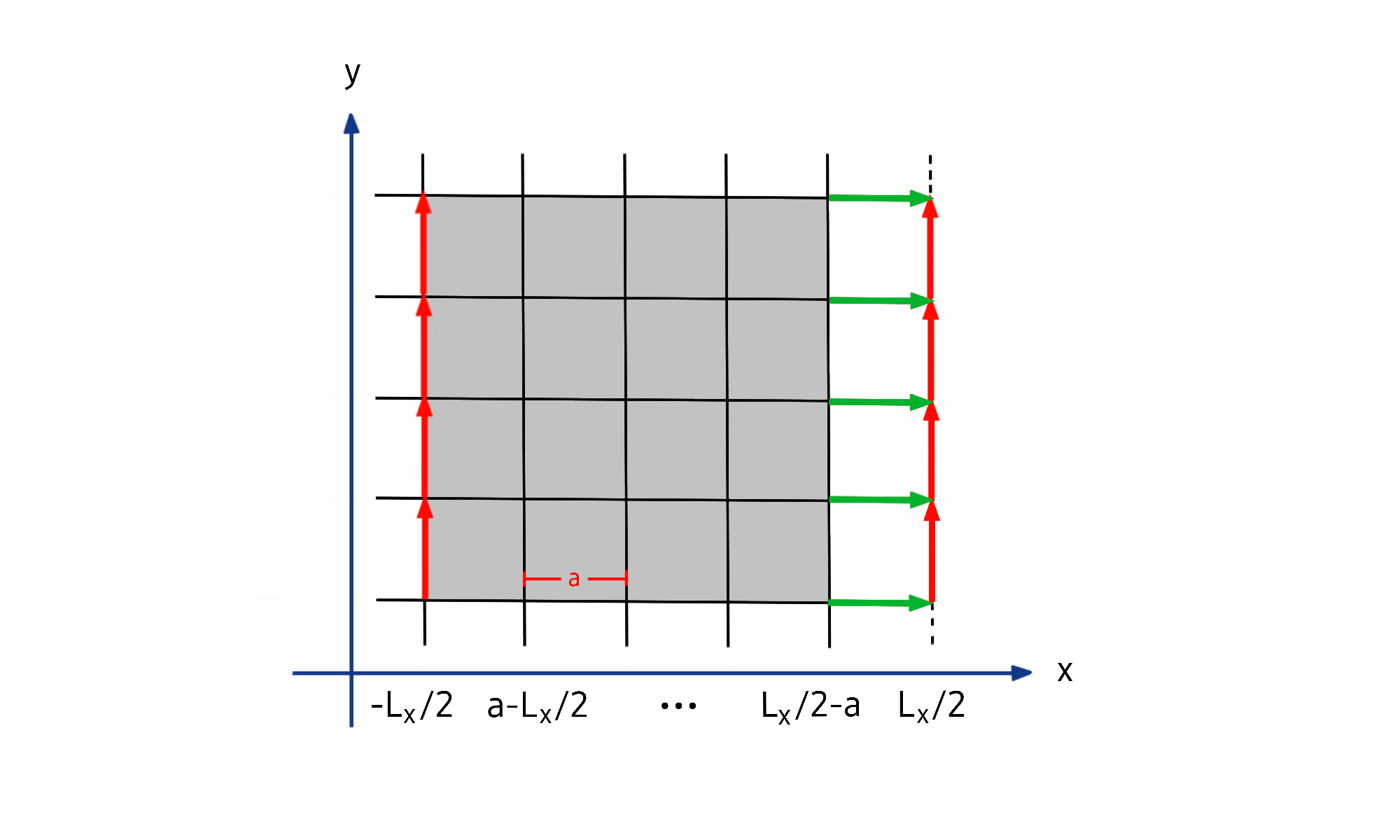}
    \caption{A gauge transformation at the last $x$-slice makes the $u_y$ links (red arrows) exactly periodic, and also affects the $u_x$ links (green arrows).}
    \label{fig:lattice_cartoon}
\end{figure}

As visualized in Fig.~\ref{fig:lattice_cartoon}, the same gauge transformation also affects $u_x(L_x/2-a,y)$,
\begin{align}
u_x(L_x/2-a,y)^{\prime} &= \Omega(L_x/2-a,y)\,u_x(L_x/2-a,y)\,\Omega(L_x/2,y)^{\dagger} \nonumber \\
&= \exp[-iq\int_{-L_y/2}^ydy'\int_{-L_x/2}^{L_x/2}dx^{\prime}\,B(x^{\prime},y^{\prime})]\,,
\end{align}
where we have used that $\Omega(L_x/2-a,y) = 1$, following the definition in Eq.~\eqref{eq:g_transform_func}. The change in the $u_x$ links ensures that the $\mathrm{U}(1)$ plaquettes are invariant under the gauge transformation, as they should. 
Finally, the prescription for the periodic links is given by
\begin{align}
u_{x}(x,y) &= 
    \left\{\begin{array}{ll}
        e^{-iq\int_{-L_y/2}^ydy'\int_{-L_x/2}^{L_x/2} dx^{\prime}\,B(x^{\prime},y^{\prime})} \quad & \text{if } x = L_x/2-a \,,\nonumber \\
        1 & \text{otherwise}\,,
        \end{array}\right. \\
u_{y}(x,y) &= e^{iaqA_y(-L_x/2,y) + iaq\int_{-L_x/2}^x dx^{\prime}B(x^{\prime},y)}\,. \label{eq:A8}
\end{align}
Now we can use this prescription and replace $B(x,y)$ by our specific profile~\eqref{eq:inv_cosh_profile}. We obtain the set of links
\begin{align}
u_{x}(x,y) &= 
    \left\{
        \begin{array}{ll}
        e^{-2iqB\epsilon (y+L_y/2)\tanh(\frac{L_x}{2\epsilon})} \qquad& \text{if } x = L_x/2-a \,,\nonumber\\
        1 & \text{otherwise}\,,
        \end{array}
    \right. \\
u_{y}(x,y) &= e^{iqB\epsilon a\qty[\tanh(\frac{x}{\epsilon}) + \tanh(\frac{L_x}{2\epsilon})]}\,.
\label{eq:links2}
\end{align}
Here we have set $A_y(-L_x/2,y)=0$, which we are allowed to do as the profile only depends on $x$.

The quantization condition for $B$ arises by enforcing the periodicity of the $u_x(L_x-a,y)$ links of~\eqref{eq:A8} in the $y$ direction. This gives
\begin{equation}
q\int_{-L_x/2}^{L_x/2} dx^{\prime} \int_{-L_y/2}^{L_y/2} dy^{\prime} \,B(x^{\prime},y^{\prime}) = 2\pi N_b, \hspace{0.5cm} N_b\in\mathbb{Z}\,,
\end{equation}
which is indeed nothing but the quantization of the total flux of the magnetic field.
For a magnetic field of the form given in Eq.~\eqref{eq:inv_cosh_profile} we obtain the quantization condition shown in Eq.~\eqref{eq:B-quantization-rule}.
Using this, together with~\eqref{eq:links2} leads to Eq.~\eqref{eq:links1} used in this work.
Note that in the limit of $\epsilon\to\infty$, these links reproduce the well-known prescription in the uniform case, as in Ref.~\cite{bali2012qcd}.

\section{Classical Dirac equation with an inhomogeneous magnetic field}
\label{app:Diracfree}

Our choice~\eqref{eq:inv_cosh_profile} is special in the sense
that (in the absence of color interactions) the Dirac equation can be solved analytically for this
profile~\cite{Cangemi:1995ee}, just like for the standard Landau problem with homogeneous magnetic fields. The lowest Landau levels are the zero modes of the two-dimensional Dirac operator (in the $x-y$ plane). Protected by the index theorem, these remain zero modes even in the inhomogeneous case. In contrast, higher Landau levels split up into $N_b$ branches, where $2\pi N_b$ is the flux of the magnetic field in a finite volume. However, besides bound states, the spectrum also contains a continuum of scattering states that are difficult to treat analytically.

\begin{figure}
    \centering
    \includegraphics[width=\textwidth]{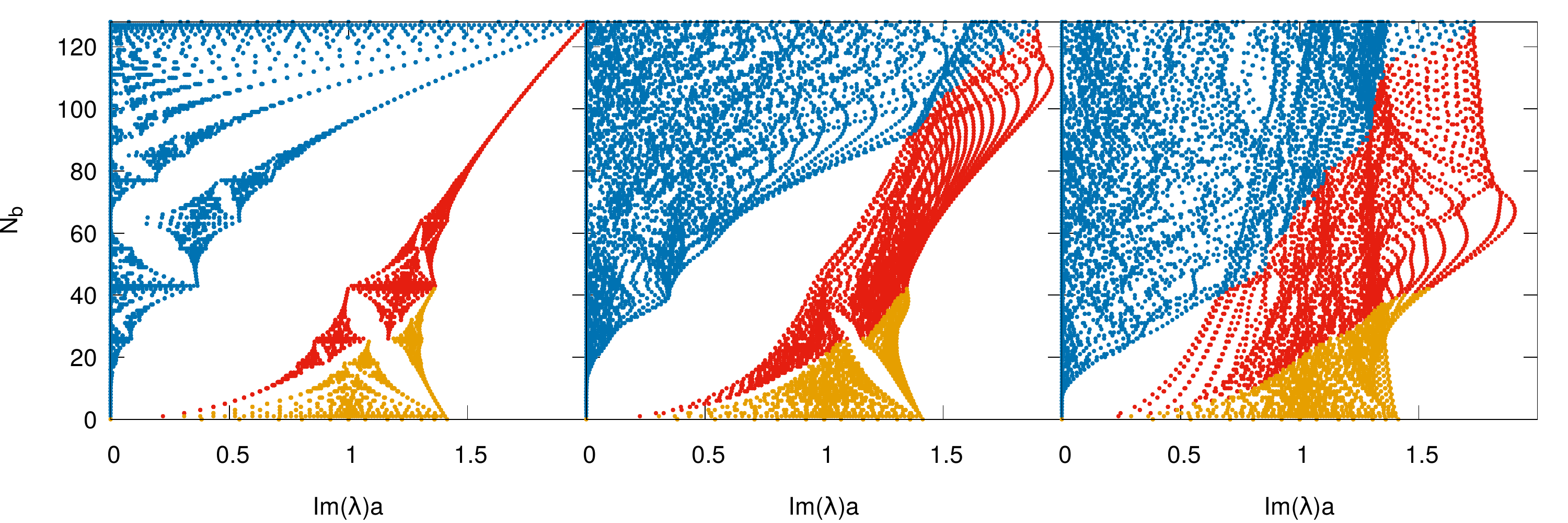}
    \caption{Eigenvalue spectrum of the staggered Dirac operator in the presence of an inhomogeneous magnetic field characterized by a flux $N_b$ and a width $\epsilon$. The three panels correspond to the homogeneous case ($\epsilon\to\infty$, left), $\epsilon=10 a$ (middle) and $\epsilon=4 a$ (right). The first $N_b$ eigenvalues (the equivalents of lowest Landau levels) are marked by blue, the next $2N_b$ eigenvalues (first Landau levels) by red, and the rest of the spectrum by orange dots. Note that the eigenvalues come in complex conjugate pairs and the plots only show the positive (imaginary) eigenvalues. }
    \label{fig:spectra}
\end{figure}

We have calculated the Dirac eigenvalues $\lambda_n$ and eigenvectors $\chi_n$ numerically on a $16^2$ lattice using the staggered discretization~\cite{Leon_thesis}. 
The lattice eigenvalue spectrum $\lambda_n(N_b)$, plotted in Fig.~\ref{fig:spectra}, constitutes a non-trivial
generalization of Hofstadter's butterfly~\cite{Hofstadter:1976zz,Endrodi:2014vza}, which is recovered 
in the $\epsilon\to\infty$ limit. Most of the recursive pattern of the homogeneous butterfly is now lost, since the magnetic field breaks translational symmetry and the spectrum is not a periodic function of the flux quantum $N_b$ anymore.\footnote{In fact, periodicity arises again for special values of the profile width, for example $\epsilon/a=1/\log(q)$ with $q>1$ a rational number and a period that is some integer multiple of $N_y$.} 
The characteristic gaps in the spectrum are also filled with eigenvalues, except for the largest gap separating lowest Landau levels from other eigenstates -- this remains a pronounced feature of the inhomogeneous butterfly due to the topological nature of the zero modes. It is interesting to point out that the same 
holds also when the homogeneous system is perturbed by color interactions~\cite{Bruckmann:2017pft}.

Using the so calculated eigenvalues, together with the chiral symmetry of the Dirac operator (for the staggered case, $\{\eta_5,\slashed{D}\}=0$ with $\eta_5=(-1)^{x+y+z+t}$), the 
local quark condensate can be reconstructed as
\begin{equation}
\bar\psi\psi(x)=\frac{1}{4}\frac{T}{L^2}\sum_{n} \sum_{p_z,p_t}\frac{m}{\lambda_n^2+p_z^2+p_t^2+m^2}
\,|\chi_n(x)|^2\,,
\end{equation}
where $p_za=\sin(2\pi n_z/N_z)$ and $p_ta=\sin(2\pi(n_t+1/2) /N_t)$ with integers $0\le n_z<N_z$ and $0\le n_t<N_t$ are the lattice momenta and fermionic Matsubara frequencies for the plane waves in the $z$ and $t$ direction, respectively, and the prefactor $1/4$ emerges due to rooting. Owing to the normalization of the modes, $\sum_x |\chi_n(x)|^2=1$, this reproduces the usual expression for the average quark condensate 
$\sum_x \bar\psi\psi(x)/L$. 

\begin{figure}
    \centering
    \mbox{
    \includegraphics[width=7cm]{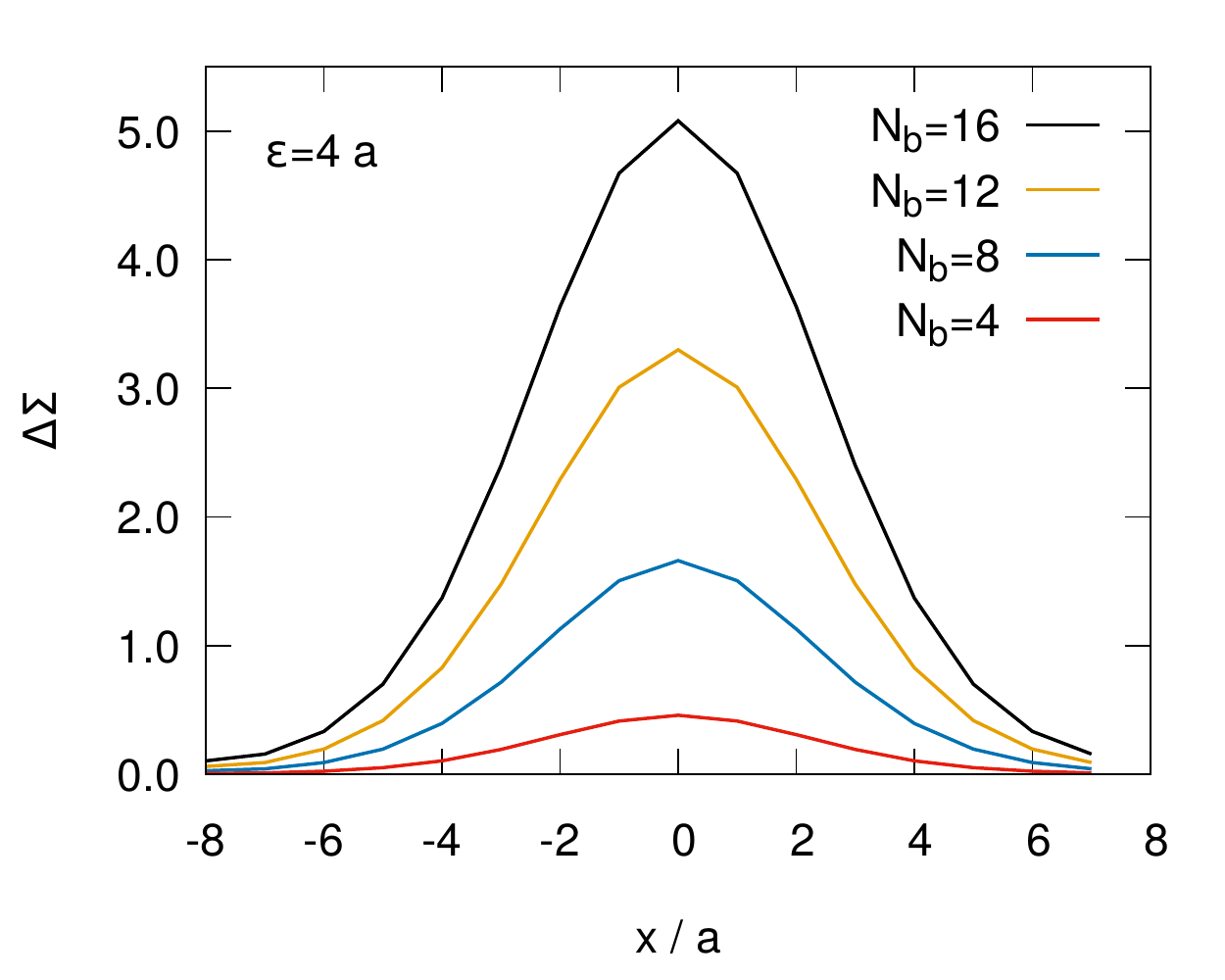} \quad
    \includegraphics[width=7cm]{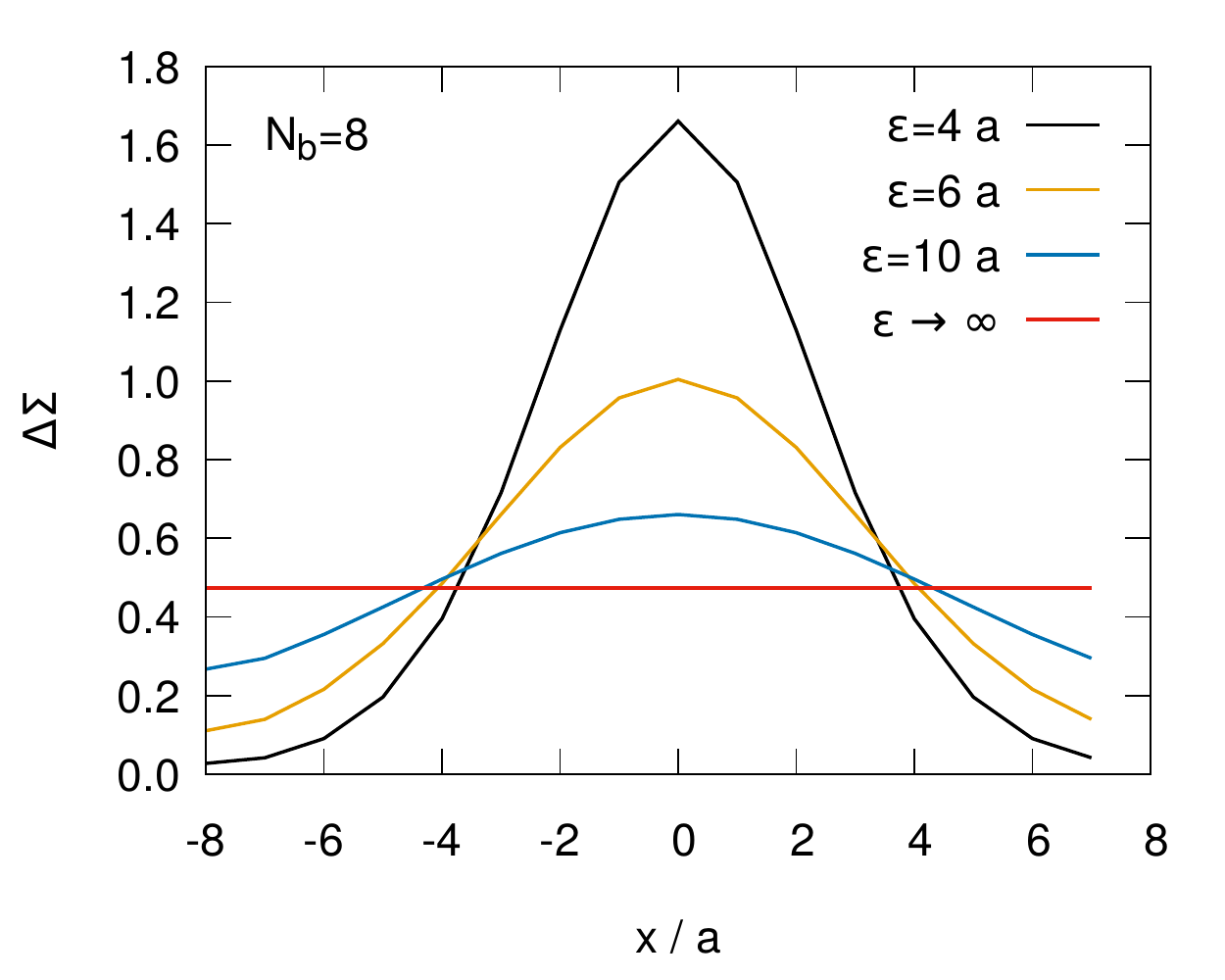}
    }
    \caption{The coordinate dependence of the subtracted condensate~\protect\eqref{eq:Sigmafdef1} in the free case for various values of $N_b$ and $\epsilon$. }
    \label{fig:freepbp}
\end{figure}

Fig.~\ref{fig:freepbp} shows
the condensate for a quark mass of $ma=0.025$ on $16^3\times4$ lattices. (Note that $m_\pi=2m$ simply in this non-interacting case but we use the physical value for $m_\pi/f_\pi$.)
The left panel of the figure clearly reveals the enhancement of the condensate 
as the amplitude of the magnetic field grows for a fixed profile. In turn, the right panel illustrates the response of the condensate for changes in the profile width. Interestingly, we find the volume-averaged condensate to be independent of $\epsilon$ as long as the lattice is sufficiently fine to resolve the profile.

\section{Classical Klein-Gordon equation with an inhomogeneous magnetic field}
\label{app:KGfree}

In contrast to App.~\ref{app:Diracfree}, which describes the high-temperature limit of QCD in terms of free quarks, here we consider the low-energy regime, i.e.\ leading-order chiral perturbation theory. The magnetic field-dependence of thermodynamic observables can be obtained by considering the one-loop determinant for a charged scalar field. The neutral pion is not affected by $B$ and thus does not contribute to observables like $\Delta\Sigma$.

\begin{figure}
    \centering
    \includegraphics[width=\textwidth]{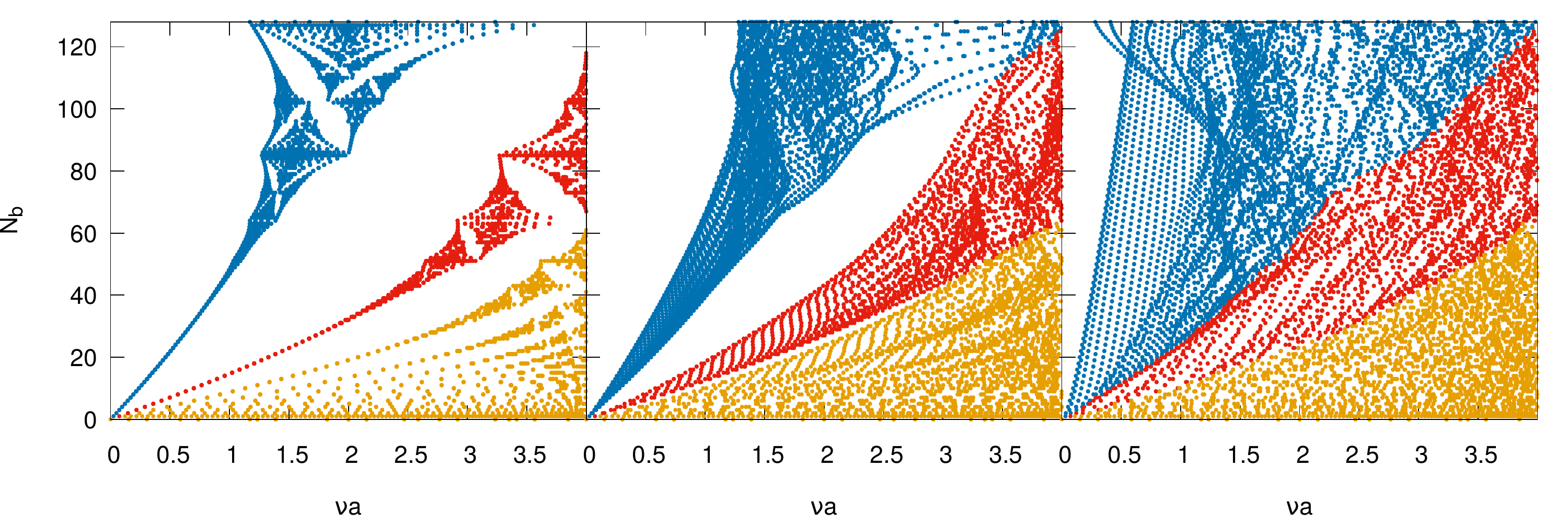}
    \caption{Eigenvalue spectrum of the Klein-Gordon operator in the presence of an inhomogeneous magnetic field characterized by a flux $N_b$ and a width $\epsilon$. The three panels correspond to the homogeneous case ($\epsilon\to\infty$, left), $\epsilon=10 a$ (middle) and $\epsilon=4 a$ (right). The first $N_b$ eigenvalues (the equivalents of lowest Landau levels) are marked by blue, the next $N_b$ eigenvalues (first Landau levels) by red, and the rest of the spectrum by orange dots. }
    \label{fig:spectra_pion}
\end{figure}

This time we need to determine the eigenvalues $\nu_n$ and eigenvectors $\varphi_n$ of the Klein-Gordon operator on two-dimensional lattices. We use the simplest symmetric discretization of the second derivative (see, e.g., Ref.~\cite{Rothe:1992nt}) including the electromagnetic parallel transporters. The spectrum $\nu_n(N_b)$ is shown in Fig.~\ref{fig:spectra_pion} for $16^2$ lattices. While for homogeneous fields, the eigenvalues can be obtained exactly from the fermionic spectrum via symmetry~\cite{Endrodi:2014vza}, this is not the case in the presence of the inhomogeneity (compare to Fig.~\ref{fig:spectra}). Contrary to the fermionic case, the lowest Landau levels are not topologically protected and thus mix with other levels as $\epsilon$ is reduced, just like higher Landau levels. Note also that due to the absence of spin, each Landau level has a degeneracy of $N_b$ and this degeneracy is lifted by the inhomogeneity.

\begin{figure}
    \centering
    \includegraphics[width=8cm]{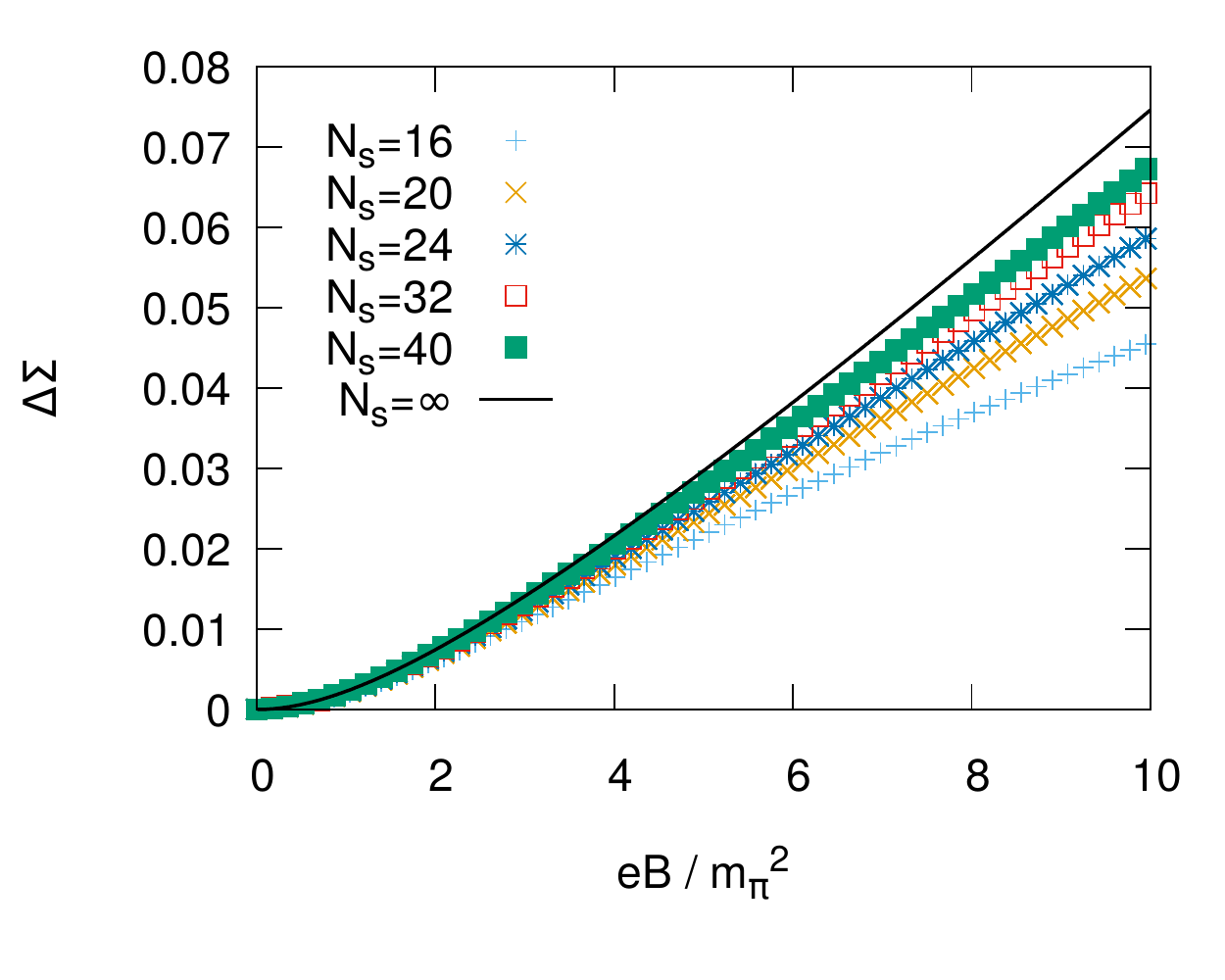}
    \caption{Numerical results for the average light quark condensate obtained using the free pion propagator for homogeneous magnetic fields on $N_s^4$ lattices, together with a comparison to standard, continuum chiral perturbation theory~\cite{Andersen:2012dz}. }
    \label{fig:chipt_comp}
\end{figure}

Using the spectrum, we calculate the traced pion propagator,
\begin{equation}
\Phi_B(x) \equiv 
-\frac{T}{L^2}
 \sum_n \sum_{p_z, p_t}
\frac{1}{\nu_n+p_z^2+p_t^2+m_\pi^2}\, |\varphi_{n_{xy}}(x)|^2\,,
\end{equation}
where the lattice momenta and Matsubara frequencies for scalar particles, $p_za=2\sin(\pi n_z/N_z)$ and $p_ta=2\sin(\pi n_t/N_t)$ with $0\le n_z < N_z$ and $0\le n_t < N_t$ enter~\cite{Rothe:1992nt}.
Using the Gell-Mann-Oakes-Renner relation $(m_u+m_d)\expval{\bar\psi\psi}=m_\pi^2F^2$, we can  trade the derivative with respect to $m_\pi^2$ for a derivative with respect to the quark masses. For degenerate quark masses, $m_u=m_d=m_{ud}$, we can
express the average light quark condensate in our normalization~\eqref{eq:Sigmadef} as
\begin{equation}
\Delta\Sigma(x) = \frac{2m_{ud}}{m_\pi^2F^2}\left[ \Phi_B(x)-\Phi_{B=0}(x) \right]
\cdot \frac{1}{2}\left[\frac{\partial m_\pi^2}{\partial m_u}+\frac{\partial m_\pi^2}{\partial m_d}\right]
= \frac{\Phi_B(x)-\Phi_{B=0}(x)}{F^2}\,.
\end{equation}
The results are checked by comparing against the well-known chiral perturbation theory formula in the homogeneous magnetic field case at zero temperature~\cite{Shushpanov:1997sf,Andersen:2012dz}. This comparison is shown in Fig.~\ref{fig:chipt_comp} for a set of $T\approx0$ lattices using $m_\pi L=6$.  The continuum limit is approached by keeping $m_\pi L$ fixed and sending $N_s=N_t\to\infty$, revealing perfect agreement with the analytical formula.

\begin{figure}
    \centering
    \mbox{
    \includegraphics[width=7cm]{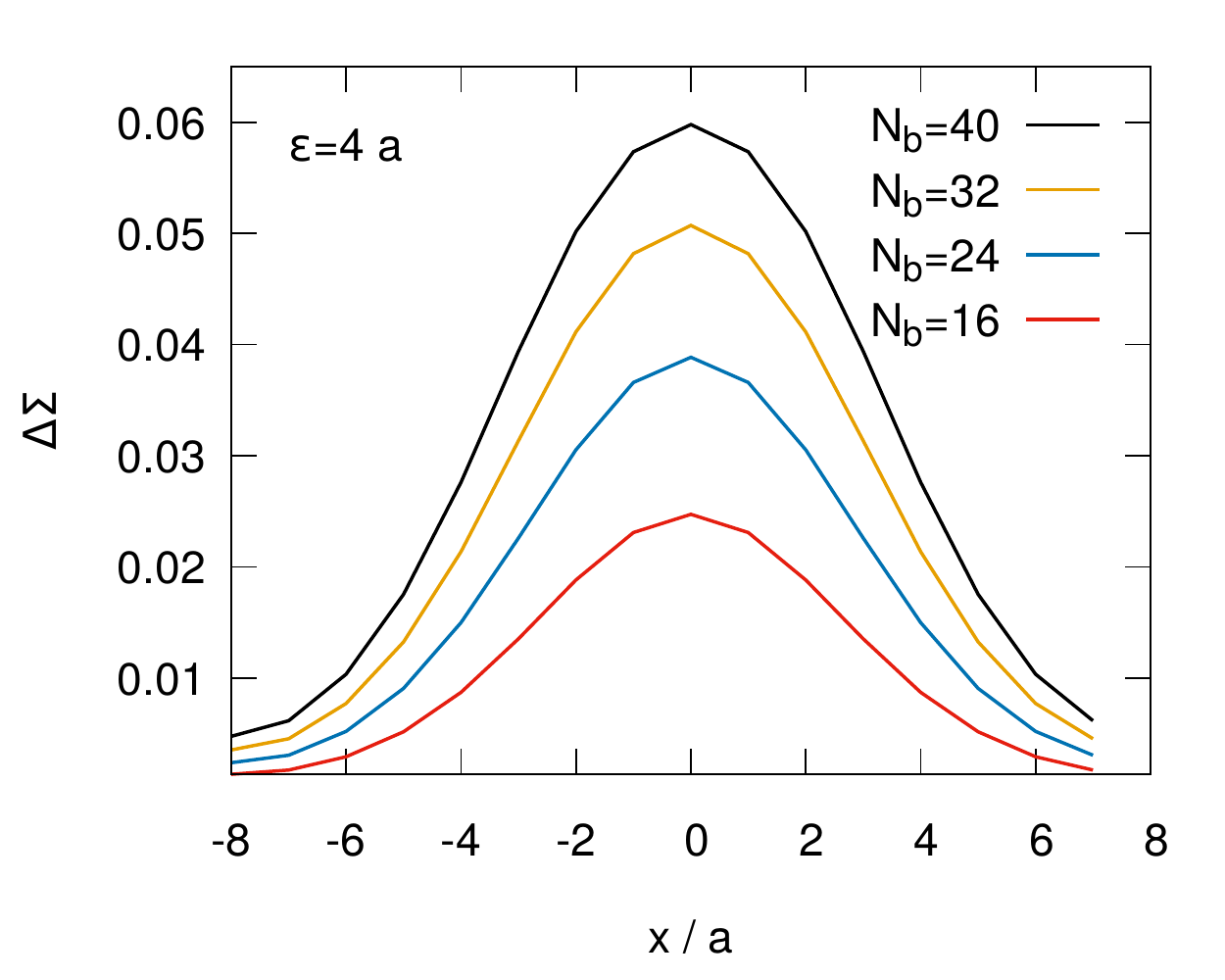} \quad
    \includegraphics[width=7cm]{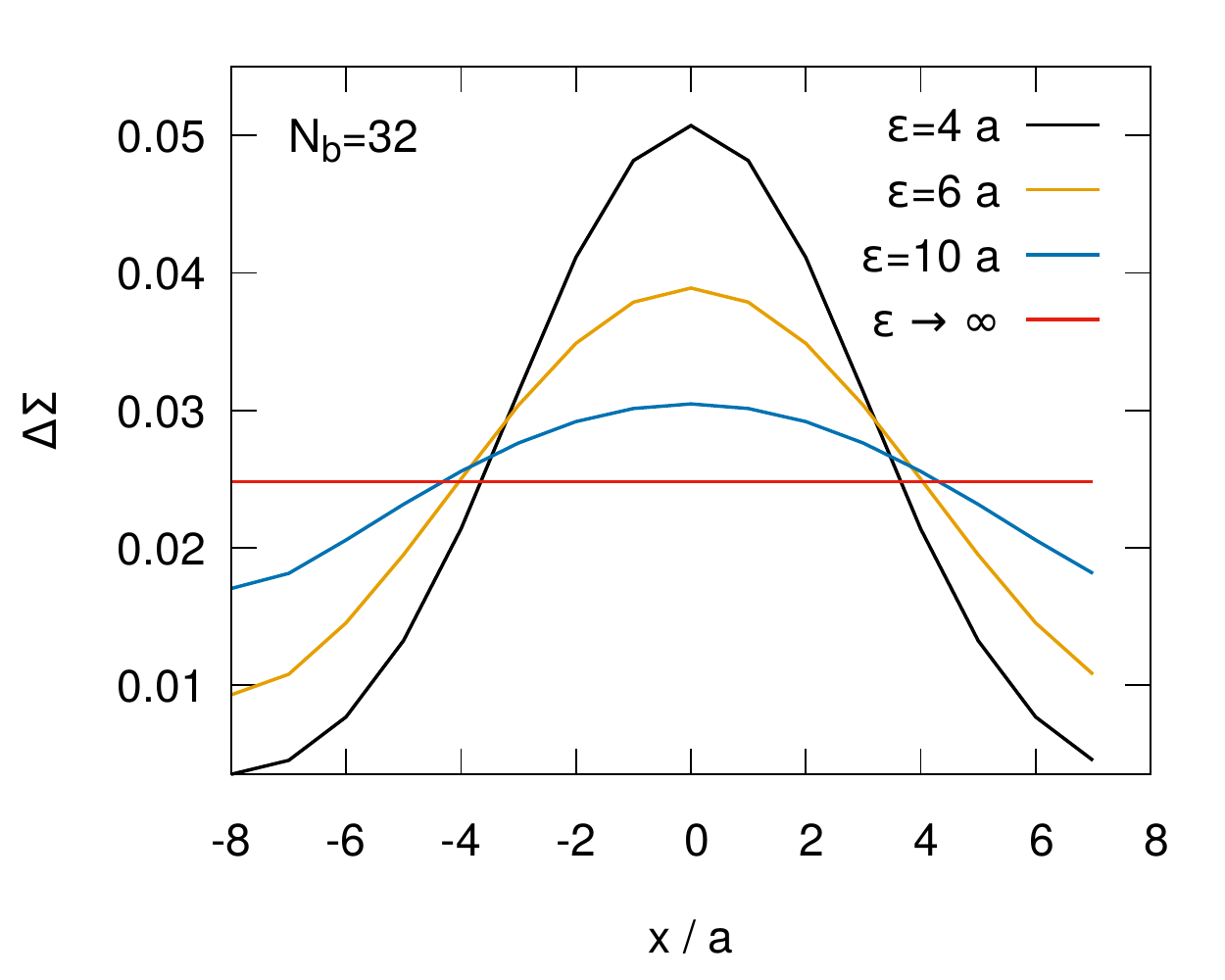}
    }
    \caption{The coordinate dependence of the subtracted condensate~\protect\eqref{eq:Sigmafdef1} for free charged pions for various values of $N_b$ and $\epsilon$. The lattice spacing is set by the pion mass, $m_\pi a=0.375$.}
    \label{fig:pionpbp}
\end{figure}

Having cross-checked our approach for homogeneous fields, we proceed to calculate $\Delta\Sigma(x)$ for inhomogeneous profiles. Fig.~\ref{fig:pionpbp} shows a scan on $16^4$ lattices for a range of widths and magnetic fluxes. 
Our results demonstrate clearly what we call `local magnetic catalysis' -- the inhomogeneous enhancement of the condensate.
Just as in the fermionic case, we again find the spatial average of the condensate to be practically independent of $\epsilon$. We compare the so obtained profiles to our full QCD results in the main body of the text.

\section{Lattice data and continuum extrapolation}
\label{app:data_continuum_limit}

In order to compute the continuum limit of our observables in the full range of magnetic fields, we need to interpolate over the discrete $N_b$ values, given by the quantization of the magnetic flux in a finite box. In this appendix, we discuss in detail our method of interpolation for the chiral condensate and the Polyakov loop.

In order to have a reliable extrapolation to the continuum and to suppress local fluctuations of the data due to statistical noise, we applied an interpolation procedure that combines a multidimensional spline fit -- defined upon a set of spline nodepoints -- and the continuum extrapolation. This is done by promoting the coefficients of the spline to be $a$-dependent and finding the best fitting surface that minimizes the `action' -- containing the $\chi^2/{\rm dof}$ as well as a term suppressing oscillatory solutions.
\newpage
\begin{figure}[!h]
    \centering
    \begin{subfigure}{0.49\textwidth}
    \includegraphics[width=\linewidth]{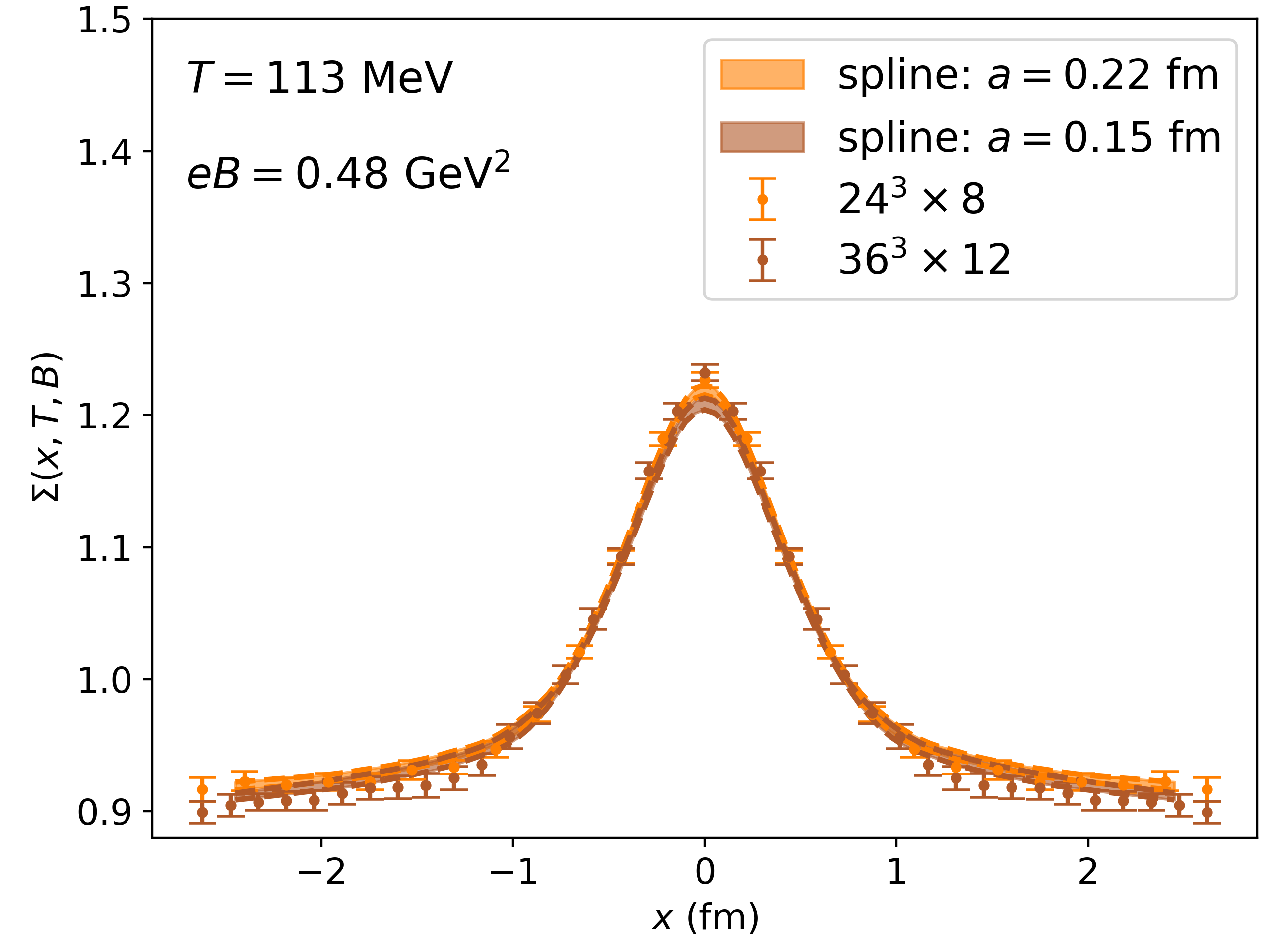}
    \end{subfigure}
    \begin{subfigure}{0.49\textwidth}
    \includegraphics[width=\linewidth]{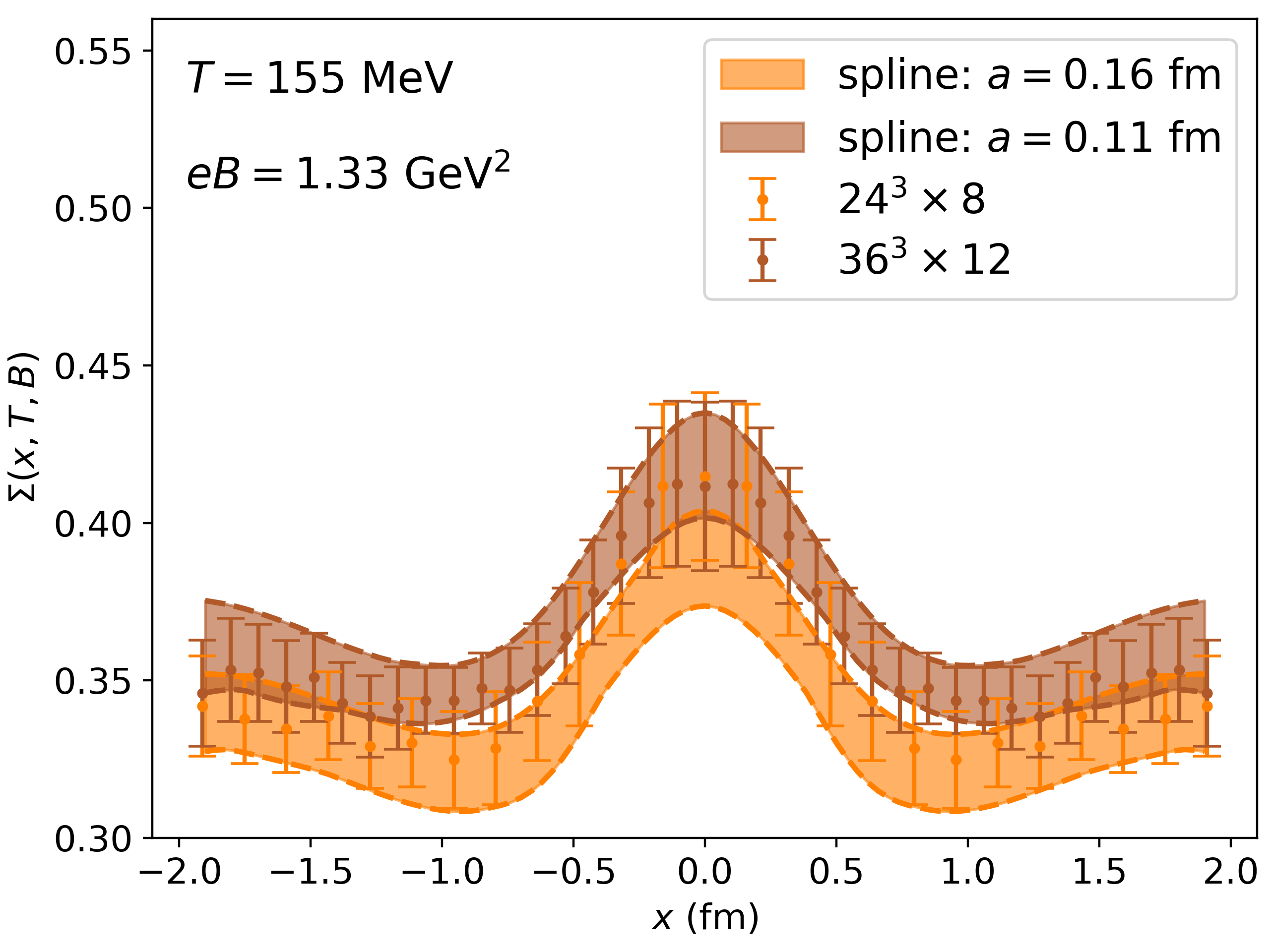}
    \end{subfigure}
    \begin{subfigure}{0.49\textwidth}
    \includegraphics[width=\linewidth]{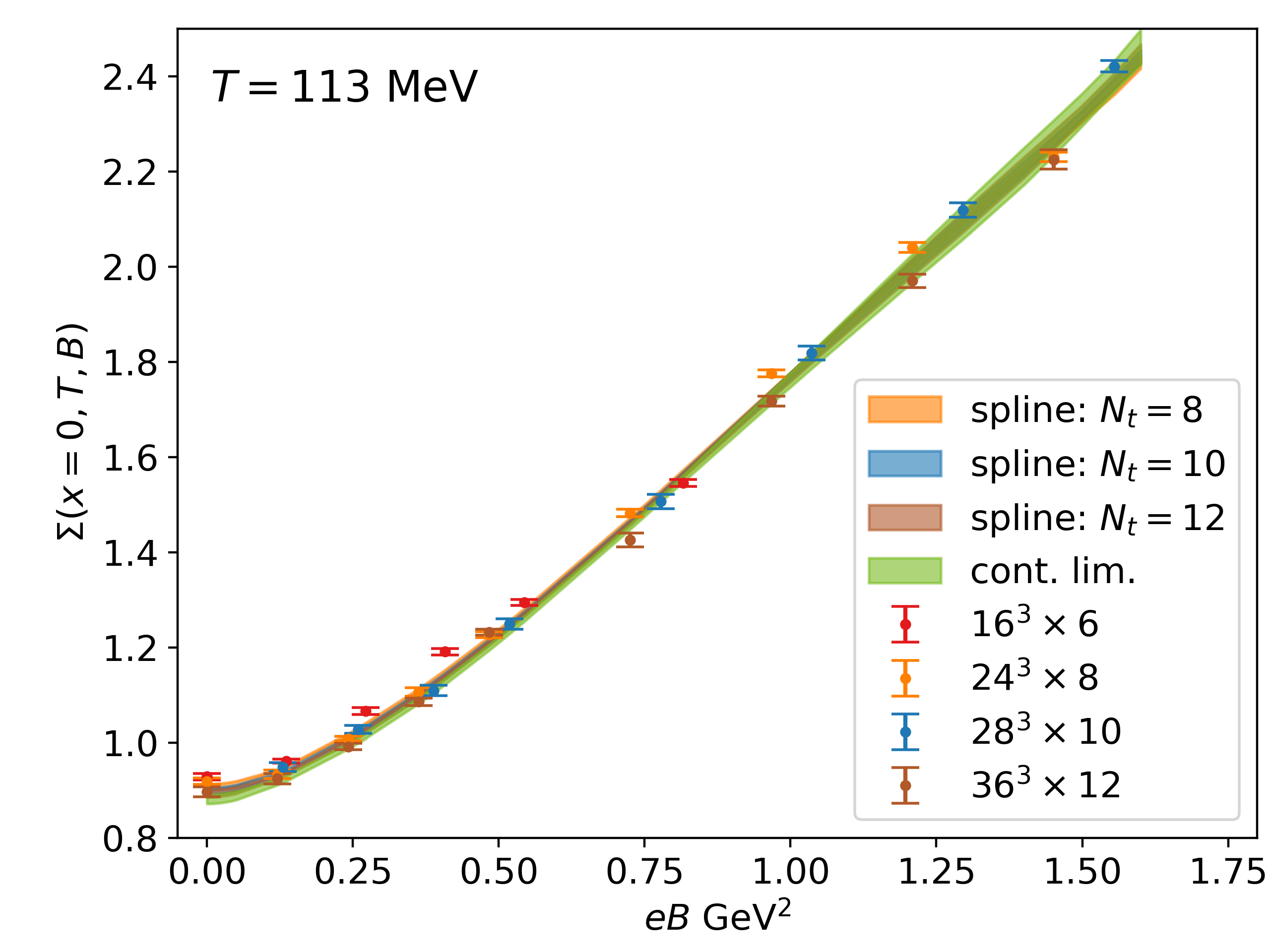}
    \end{subfigure}
    \begin{subfigure}{0.49\textwidth}
    \includegraphics[width=\linewidth]{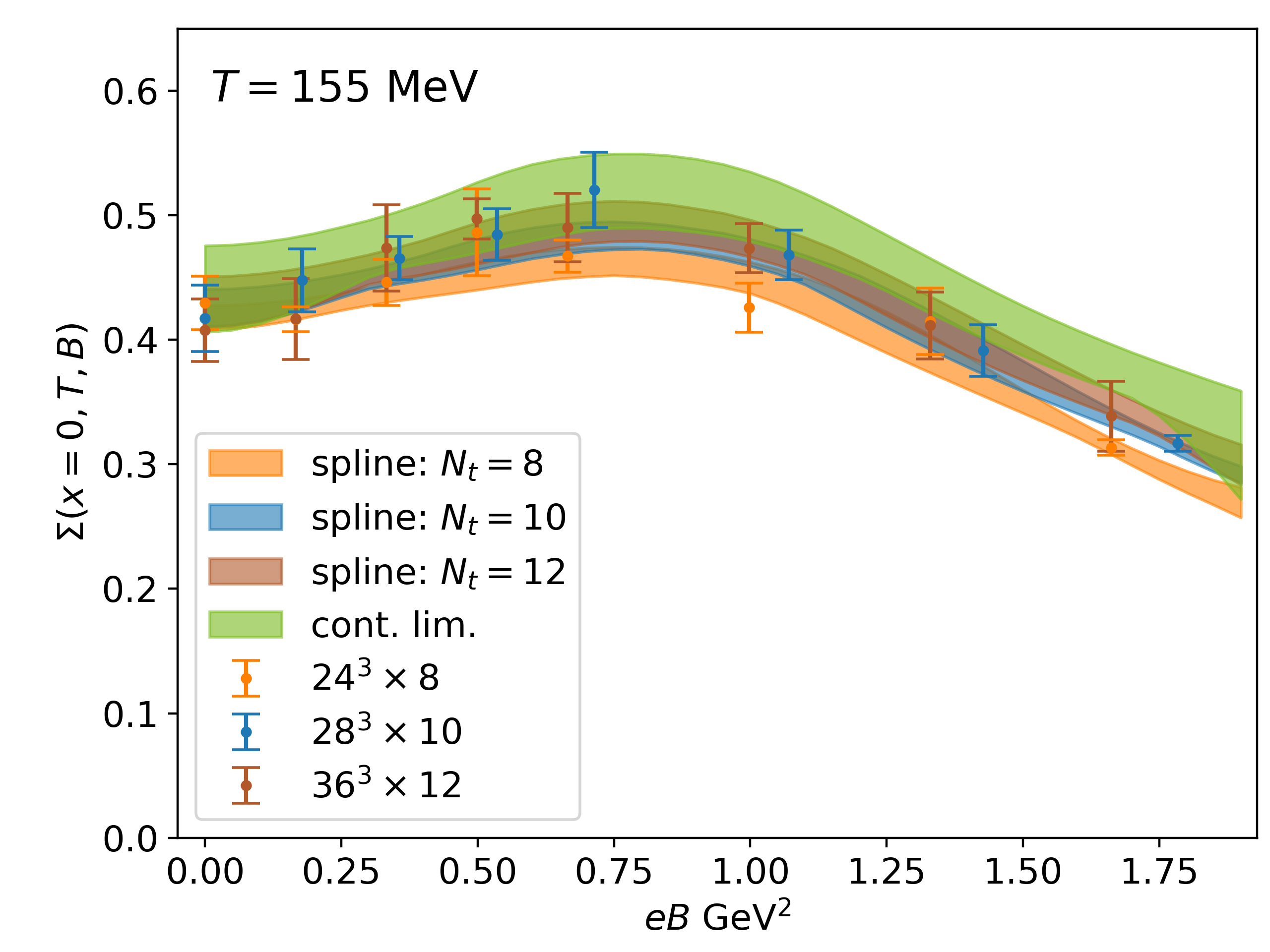}
    \end{subfigure}
    \begin{subfigure}{0.49\textwidth}
    \includegraphics[width=\linewidth]{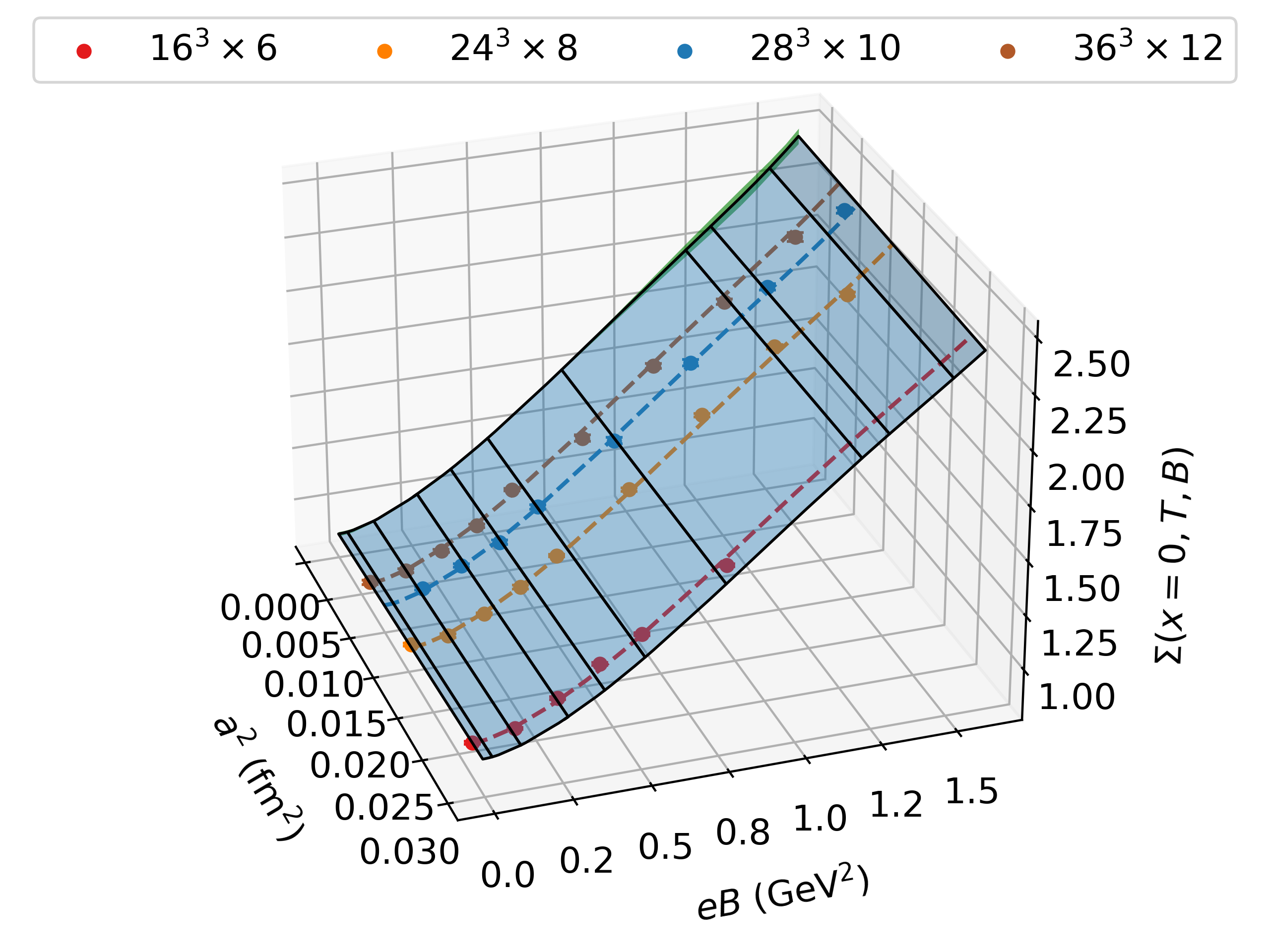}
    \end{subfigure}
    \begin{subfigure}{0.49\textwidth}
    \includegraphics[width=\linewidth]{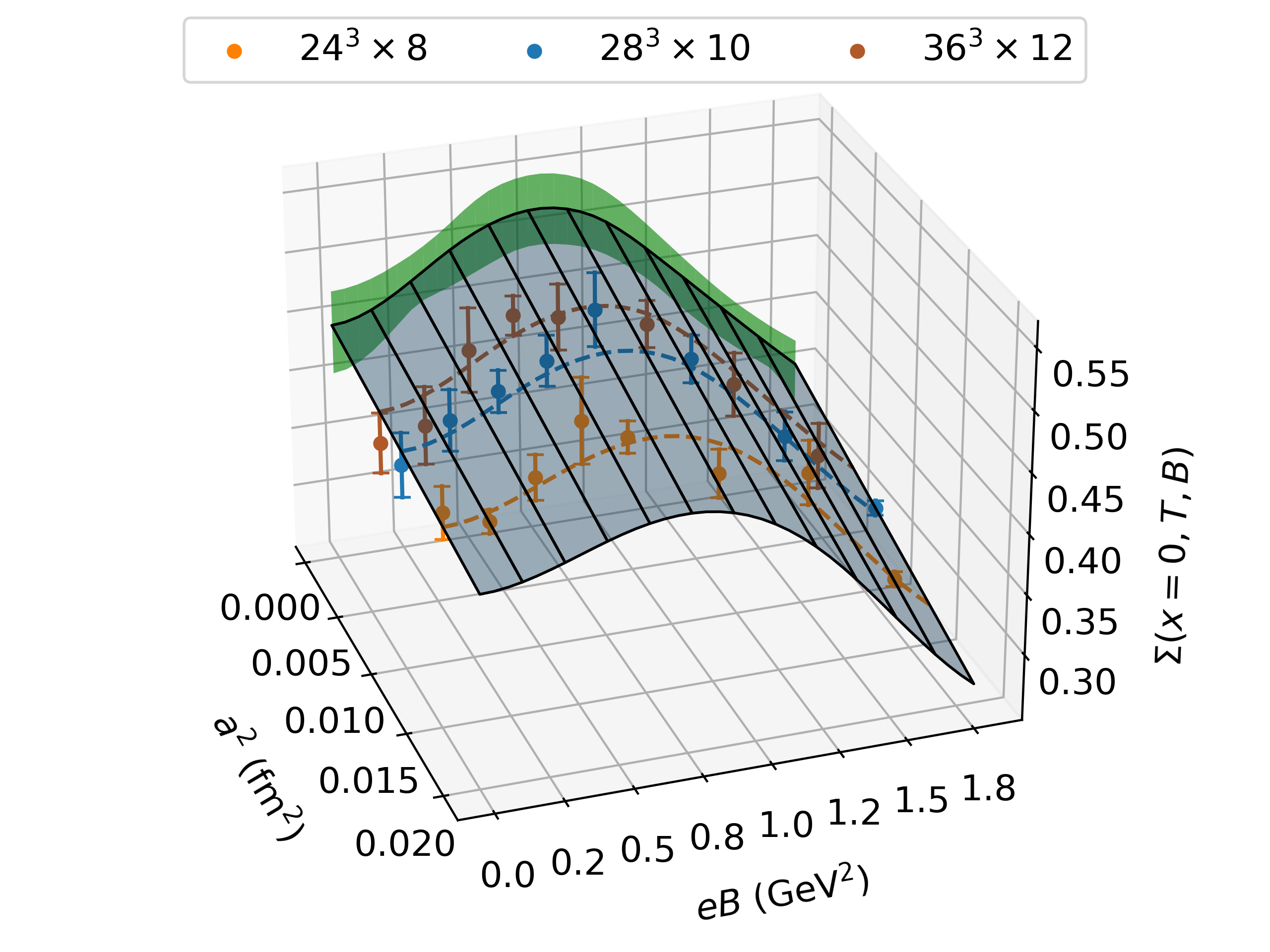}
    \end{subfigure}
    \caption{Top plots: chiral condensate as a function of $x$ for different $T$. The lattice results (data points) are compared to the one-dimensional slice of the spline fit at the corresponding values of $a$ and $eB$. Middle plots: chiral condensate at $x = 0$ as a function of $eB$ obtained on different lattices (data points) and the continuum limit extracted from the spline fit (green band). Bottom plots: $a$-scaling of the spline fits at $x = 0$ for different $eB$ with continuum limit projection (green band). The error bands include statistical and systematic errors added in quadrature.}
    \label{fig:spline_interp_cond_T=113}
\end{figure}
\newpage
To assess systematic errors, a Monte-Carlo procedure is set up to generate fits having different numbers and positions of spline nodepoints, according to this action. The details of the spline fit and the Monte-Carlo algorithm are explained in Refs.~\cite{Endrodi:2010ai} and~\cite{Brandt:2016zdy,Brandt:2022hwy}, respectively.

In particular, at each $T$, we considered the two-dimensional space spanned by $x$ and $eB$ including a lattice spacing dependence, and performed a series of Monte Carlo updates for the corresponding spline fits.
Owing to the scaling properties of our lattice action, the lattice spacing-dependence of the spline was fixed to be quadratic and we included the leading $\mathcal{O}(a^2)$ lattice artefact term.
To constrain our spline fits, we exploited parity symmetry, i.e.\ that the observables are symmetric in $x$ around $x = 0$ and in $B$ around $B=0$.Thus, we imposed the conditions
\begin{equation}
\pdv{\Sigma(x,B;a)}{x}\Bigg|_{x = 0} = 0 \hspace{1cm} \pdv{\Sigma(x,B;a)}{(eB)}\Bigg|_{eB = 0} = 0\,,
\end{equation}
and similarly for the Polyakov loop.
We chose the initial position of the spline nodepoints individually for each $T$, given that the observables, for instance the quark condensate, have various degrees of complexity with $x$ for different $T$.
The systematic errors are accounted for by averaging over different possible spline solutions, generated from the Monte-Carlo algorithm.
Note that the $B=0$ slice of the splines is not constrained to give an $x$-independent constant -- they are observed to fluctuate around the $B=0$ value of the observables. 
In Fig.~\ref{fig:spline_interp_cond_T=113}, we show the results of our spline fit for the chiral condensate at a selected set of parameters. To avoid enhanced lattice discretization effects, in the case of the chiral condensate we excluded our $16^3\times6$ data at $T\gtrsim 140 \textmd{ MeV}$ from the global fit.
\bibliographystyle{JHEP}
\bibliography{bibliography.bib}
\end{document}